\documentclass{emulateapj}  
\usepackage{latexsym}  
\newcommand{\beq}{\begin{equation}}
\newcommand{\eeq}{\end{equation}}

\begin{document}

\shorttitle{Statistics of Dark Matter Clusters}
\shortauthors{Shaw et al.}
\title{Statistics of Physical Properties of Dark Matter Clusters}

\author{Laurie D Shaw\altaffilmark{1}, Jochen
Weller\altaffilmark{2,3},Jeremiah P Ostriker\altaffilmark{1,4},Paul
Bode\altaffilmark{4}}
\affil{${}^1$Institute of Astronomy, University of Cambridge,
  Madingley Road, Cambridge CB3 0HA, UK. \\ ${}^2$NASA/Fermilab Astrophysics Group, Fermi National
  Accelerator Laboratory, Batavia, IL 60510-0500, USA. \\ ${}^3$Department of Physics and Astronomy, University College
  London, Gower Street, London WC1E 6BT, UK. \\ ${}^4$Princeton University Observatory,
  Princeton NJ 08544-1001, USA.}

\begin{abstract}

  We have identified over 2000 well resolved cluster halos, and also
  their associated bound subhalos, from the output of a $1024^3$
  particle cosmological N-body simulation (of box size 320$h^{-1}$Mpc
  and softening length 3.2$h^{-1}$kpc). This has allowed us to measure
  halo quantities in a statistically meaningful way, and for the first
  time analyse their distribution for a large and well resolved
  sample. We characterize each halo in terms of its morphology,
  concentration, spin, circular velocity and the fraction of their
  mass in substructure. We also identify those halos that have not yet
  reached a state of dynamical equilibrium using the virial theorem
  with an additional correction to account for the surface pressure at
  the boundary. These amount to 3.4\% of our initial sample.  For the
  virialized halos, we find a median of 5.6\% of halo mass is
  contained within substructure, with the distribution ranging between
  no identified subhalos to 65\%. The fraction of mass in substructure
  increases with halo mass with logarithmic slope of $0.44 \pm 0.06$.
  Halos tend to have a prolate morphology, becoming more so with
  increasing mass. Subhalos have a greater orbital angular momentum
  per unit mass than their host halo. Furthermore, their orbital
  angular momentum is typically well aligned with that of their
  host. Overall, we find that dimensionless properties of dark matter
  halos do depend on their mass, thereby demonstrating a lack of
  self-similarity.
 
\end{abstract}

\section{Introduction}

N-body dark matter simulations have proven to be one of our most
valuable tools for testing the predictions of the standard model of
hierarchical structure formation. Indeed, it is largely through
studying the structures that form in such simulations that we are able
to investigate proposed models that describe the relative mass
distribution of halos and the subhalos within them, as well as their
physical characteristics such as spin, shape, and concentration.

Until the end of the last decade, it was not possible reach the
required mass resolution to enable the identification of galaxy mass
halos as substructures in clusters \citep{White:76, vanKampen:95,
Summers:95, Moore:96}.  Commonly known as the {\it overmerging}
problem, this was mainly due to the limited mass and force resolution
of the simulations used.  The major causes of this problem were
premature tidal disruption due to inadequate force resolution and
two-particle evaporation for halos with a small number of particles
\citep{Klypin:99}. However, rapid advances in parallel computing,
through both the improvement of hardware and the development of fast
and efficient parallel algorithms, has enabled us to achieve the
numerical resolution required to overcome the numerical problems
\citep{Ghigna:98, Klypin:99, Moore:99a, Okamoto:99, Ghigna:00,
Springel:01a, Bode:01, Kravtsov:04b, DeLucia:04, Springel:05b}. 

There are two preferred approaches for investigating the physical
properties of simulated dark matter halos. In order to achieve high
numerical precision--- thus allowing the accurate identification of
low mass subhalos and measurement of the properties in the innermost
regions--- many previous studies have adopted a resimulation technique
(e.g. \cite{Navarro:95}, \cite{Navarro:96}, (hereafter referred to as
NFW), \cite{Gao:05}, \cite{Navarro:04}, \cite{Moore:99a},
\cite{Reed:04}). This involves extracting a halo from a cosmological
simulation and resimulating it at much higher resolution,
approximating the tidal field of the surrounding structures.  By
picking a `nice' (or relaxed) halo in the original simulation, it can
furthermore be ensured that the final halo will be in dynamic
equilibrium. Using this technique, \citet{Diemand:05} completed the
first billion particle halo simulation, achieving unprecedented
resolution and probing the halo density profile to within 0.1$\%$ of
the halo virial radius.

However, there are two downsides to this approach. Firstly,
resimulating a small number of halos prevents an accurate
determination of the {\it distribution} their physical
properties. Previous studies have typically resimulated several halos
in an attempt to ascertain if geometric and dynamical properties
differ with mass.  There remains a significant degree of uncertainty
in the results of such studies merely because of the small number
statistics involved.  The resimulated halos do not present an accurate
picture of the entire range of physical characteristics that may be
adopted by dark matter halos when one chooses only relaxed halos for
higher resolution.

The second approach that one can take is to uniformly simulate a
cosmological volume at the highest resolution possible (given the
available resources) and pick out a large number of halos. Although
the ability to achieve higher possible mass resolution is lost, this
method results in a statistically significant sample of halos. A
number of previous studies (e.g. \cite{Davis:85}, \cite{Warren:92a},
\cite{Cole:96}, \cite{Bullock:01b}) employed this approach to analyse
the distribution of the integral properties of a large sample of
halos, including their virial mass, spin and overall distribution over
the volume simulated; or they used the dark matter distributions to
estimate the large angle gravitational lensing of background sources
\citep[e.g.][]{Wambsganss:04}. \citet{Kravtsov:04b} performed
simulations with high enough resolution to identify substructures and
even subsubstructures within the cluster halos extracted from the
simulation box. They analyze the halo occupation distribution for the
host halos and the two-point correlation function of the galaxy size
halos and subhalos, finding the results to be in good agreement with
those from the Sloan Digital Sky Survey. Recently,
\citet{Springel:05b} presented initial results from the {\it
``Millenium''} simulation -- the highest resolulation cosmological
N-body simulation to date -- containing $2160^3$ particles in a cube
of side length $500 h^{-1} Mpc$. This simulation was performed using
the new parallel GADGET2 code \citep{Springel:05} which adopts a
variant of the Tree-Particle-Mesh (hereafter, TPM) technique for
calculating the gravitational forces on each particle \citep{Bode:01,
Bode:03a}.

This paper takes the second approach, presenting an analysis of a
$1024^3$ particle simulation with box size 320 $h^{-1}$ Mpc evolved
using the TPM algorithm of \citet{Bode:03a} and containing more than
57,000 structures at the last timestep. Of these we have analysed the
most massive 2200 cluster halos containing at least 10,000 particles
each, thus achieving unprecedented resolution in such a large sample
of halos. The clusters are first identified using a friends-of-friends
algorithm; we then employ the geometrically based Denmax routine of
\citet{Bertschinger:91} to identify the subhalos in each cluster and
build family trees.  Finally, all energetically unbound particles are
removed. The algorithm we employ up to this stage is described in
detail in \citet{Weller:04}.

We first present a procedure to identify the halos that are still in
the processes of formation and have not yet achieved a dynamically
relaxed, virialized state. The purpose of this is to enable an
independent analysis of the physical characteristics of those halos
that are dynamically relaxed. A catalogue of the main physical
properties of the halos is then compiled, including morphology, spin,
fraction of mass in substructure, the subhalo mass function,
concentration and maximum circular velocity. We analyse the
distributions of each quantity, which are well defined due to the size
of our sample, the dependence, if any, on halo mass, and how halo
substructure influences overall cluster properties. The key questions
we ask therefore, are
\begin{itemize}
\item{What is the effect of the dynamical state of the halo on the
physical properties that we measure?}
\item{What are range the of values that the physical properties of the
halos take (in the mass range we investigate)?}
\item{Are cluster halos self-similar, i.e. how do the properties of halos vary with halo mass?}
\item{What is the effect of substructure on the overall characteristics of a cluster?}
\end{itemize}

In Section 2 we describe in detail the sample of halos that we analyse
in this paper and how unvirialized clusters can be identified. In
Section 3 we discuss the physical properties that we measure and how
we do so. In Sections 4, 5, and 6 we present the main results of our
analysis--- the distribution of halo properties, how they depend on
halo mass and the impact of substructure on halo properties. Finally,
in Section 7 we discuss the results and state our conclusions.

\section{Definition of Sample}
\label{sec:defsam}
\subsection{The Identification of Halos and Subhalos}

There are many possible algorithms for defining and identifying halos
and subhalos, and there is no generally agreed upon best method
adopted by current investigations.  Some methods are essentially
geometrical and others aim at finding dynamically coherent structures;
all contain both dimensionless and dimensional parameters.  In
general, there are three levels of refinement that one can adopt. The
most basic is to use a geometrical routine such as the
friends-of-friends \citep{Huchra:82, Davis:85, Lacey:94} or Denmax
\citep{Bertschinger:91, Gelb:94, Eisenstein:98} algorithms.  These use
only instantaneous particle positions to group together nearby
particles (defined by the FOF `linking length' or the Denmax
`smoothing length') into localized structures. Neither method performs
any type of dynamical analysis to check whether these structures are
bound.

The next level of refinement is to follow a geometrical identification
of subhalos with a procedure for iteratively removing the unbound
particles with the greatest total energy until only bound particles
remain. Examples of routines that use this approach include SKID
\citep{Stadel:97a}, BDM \citep{Klypin:99}, MHF \citep{Gill:04a} and
SUBFIND \citep{Springel:01a}.  In each of these, the unbinding is
performed assuming the subhalo is completely isolated, i.e. only the
bound particles in the halo at each iteration are taken into
consideration when calculating the potential energy of each
particle. Not taken into account in this energy calculation are the
(previously identified) unbound particles located spatially within the
subhalo, nor the disruptive effect of the tidal forces from the
particles surrounding it. In \cite{Shaw:06} we investigate the impact
of allowing for these forces, finding there to be little difference in
the subhalo populations measured; the increase in the binding
energy of a subhalo from including all the particles located within it
is almost entirely balanced by the losses due to the external
forces.

In this paper we follow previous authors and implement only the first
refinement--- identifying energetically bound structures--- to an
initial geometrical selection of structures. The algorithm is
described in detail by \citet{Weller:04} and is applied to a
$\Lambda$CDM N-body simulation of $1024^3$ particles with box size
$320 h^{-1}\,{\rm Mpc}$ and a spline kernel force softening length
\citep{Hernquist:89} of $\epsilon = 3.2h^{-1}{\rm kpc}$.  This
simulation was evolved to $z$=0.05 using the Tree-Particle-Mesh (TPM)
algorithm \citep{Bode:03a}.  The cosmological parameters used include
$\Omega_m=0.3$, $\Lambda=0.7$, and $\sigma_8$=0.95; outputs from this
run have previously been used to make predictions concerning strong
lensing \citep{Wambsganss:04, Hennawi:05}. We wish to ensure that all
the halos are well resolved and that the overmerging problem is not in
evidence.  Thus we discard all halos with less than 10,000 particles.
The minimum halo mass allowed is therefore $\approx 3\times 10^{13}
h^{-1} M_{\rm \sun}$.

The method described in \citet{Weller:04} starts with the hierarchical
identification of structures: firstly the large clusters are
identified through a Friends-of-Friends (FOF) routine with a linking
length of $b=0.2\bar{n}^{-1/3}$, where $\bar{n}^{-1/3}$ is the mean
inter-particle separation. The Denmax routine of
\citet{Bertschinger:91} is then run on each FOF halo with a high
resolution smoothing length of $5\epsilon$ in order to identify the
substructures within each cluster halo.  A family tree is then
constructed by hierarchically associating the smallest mass subhalos
with their lowest mass `ancestor', so that each subhalo has only a
single immediate parent. Those that consist of less than 30 particles
are dissolved into their immediate parent.

Up until this point the analysis is purely geometrical. Next,
energetically unbound particles are discarded, first by removing the
velocity outliers, and then by iteratively identifying those particles
with a total energy greater than zero (in the center of mass frame of
the halo) and removing the most energetic. If at any stage the mass of
a subhalo drops below 30 particles, we dissolve it into its immediate
parent. We then check to see if pairs of the immediate daughters of
the parent cluster halo are bound. If this is the case, they form a
hyperstructure: the less massive of the two becomes a daughter of the
more massive structure. Finally, we remove all subhalos that are not
bound to the biggest structure--- the mother halo--- and any particles
that are not bound to the entire cluster. As described in
\citet{Weller:04}, this entire procedure is both stable and largely
independent of arbitrary parameters choice. Henceforth, we refer to
the most massive structure in each halo as the `mother' or `host'
halo. Substructure particles are those that are also bound to any
other smaller structure within the halo. We use `cluster' or `{\it the
halo}' to refer to the entire assembly.

In this study, we also incorporate an additional step: the tagging of
`irregular' halos in our sample. The purpose of this is to allow a
separate analysis of those halos that are still undergoing significant
changes through their involvement in a major-merger. This is described
in the following Section.

\subsection{Virialization Criterion--- Identifying Irregular Halos}
\label{sec:virial}

One of the main aims of this study is to compare the physical
properties of the virialized halos in our sample to those that
are still in the process of forming. To achieve this we sort through
the halo catalogue to isolate and tag two types of halo:

\begin{itemize}
\item{those with mass below a certain threshold, to ensure that all
the halos are well resolved and that the overmerging problem is not in
evidence. These are eliminated from the sample.}
\item{halos that we deem to have not yet reached a state of dynamical
equilibrium. These are tagged so that we can analyse their physical
characteristics separately.}
\end{itemize}

To satisfy the first condition, we discard all halos with less than
10,000 particles, or $\approx 3\times 10^{13} h^{-1} M_{\rm \sun}$. We
use the virial mass as our measure of halo mass. For a $\Lambda$CDM
cosmology, it is conventional to define the virial mass $M_{\rm vir}$
and radius $R_{\rm vir}$ as $M_{\rm vir}= \frac{4}{3}\pi R_{\rm vir}^3
\Delta_c(z) \rho_c(z)$, where $\rho_c$ is the critical density of the
universe, and the mean over-density $\Delta_c=178\Omega_{\rm
m}(z)^{0.45}$ \citep{Lahav:91}. In order to calculate the virial mass
of each halo, we start at its density maximum and proceed outward
until we reach the virial radius, within which the mean over-density
is $\Delta_c$. Further, we include substructures whose center of mass
is within the virial radius. Hence we, define a particle as being part
of the halo if it is bound to the halo and lies within the virial
radius, or is part of a bound subhalo that has its center of mass
within the virial radius. When quoting halo masses, we
do so relative to the characteristic mass scale, $M_* = 8.0 \times
10^{14} h^{-1} M_{\rm \sun}$, obtained by fitting a Schechter function
to the mass distribution of all halos with greater than 5000 particles
in the simulation box (see Figure 2 in \cite{Weller:04}). At this
stage, having removed the low mass halos, we have an overall sample of
2159 halos.

For the second requirement, we use the virial theorem to determine
which halos are not dynamically relaxed. The virial theorem for a
system in equilibrium is usually stated as $2T_{0} + W_{0} = 0$, where
$T_{0}$ is the total kinetic energy of the halo, and $W_{0}$ the total
potential energy. However, this assumes a system with finite phase
space volume, and also that each halo is in complete isolation and
that all mass associated with the halo has been considered. In
selecting a halo, a cut is made at some outer boundary, in this case
corresponding to the isodensity contour at which the spherically
averaged density is $\Delta_c \approx 100$ times the critical density
\citep{Lahav:91}.  Any particles bound to the halo outside of the
virial radius are by the standard convention not included when we
calculate the total gravitational and kinetic energy for each halo.
However, these particles do make a significant overall contribution to
the pressure at the halo boundary, and so we must incorporate an
additional term into the virial theorem to account for them. This
correction manifests itself as a form of surface pressure at the
boundary \citep{Chandrasekhar:61} of the halo, with the energy content
\begin{equation}
E_{s} = \int P_s(r) \mathbf{r} \cdot \mathbf{dS}\; .
\label{eq:Es}
\end{equation}

The full and exact version of the virial theorem for a
self-gravitating system is thus
\begin{equation}
\frac{1}{2}\frac{d^2I}{dt^2} = 2T_{0} + W_0 - E_s\; ,
\label{eq:fullvirial}
\end{equation}
where I is the moment of inertia of the halo. We therefore define
as our measure of `virialization' the following:
\begin{equation}
\beta \equiv \frac{2T_{0} - E_s}{W_0} + 1\; ,
\label{eq:beta}
\end{equation}
where $\beta \rightarrow 0$ as $\ddot{I} \rightarrow 0$.

In practice, we compute $P_{s}$ in the following manner: First, we
rank order all the particles in the halo by radius and select the
outermost 20$\%$. We label the radius of the innermost particle in
this shell as $R_{0.8}$, the outermost as $R_{\rm vir}$ and the median as
$R_{0.9}$.
Using the ideal gas law, we can approximate the surface
pressure term,
\begin{equation}
P_{s} = (1/3)\frac{\sum_{i}(m_i v_i^2)}{V} \;, 
\label{eq:Ps}
\end{equation}
summing over all particles that lie between $R_{0.8}$ and $R_{\rm vir}$. V is
the volume occupied by the outermost 20$\%$,
\begin{equation}
V = \frac{4\pi R_{v}^{3}}{3} - \frac{4\pi R_{0.8}^{3}}{3}\; ,
\end{equation}
hence Equation \ref{eq:Es} can be approximated by
\begin{equation}
E_{s} \approx 4\pi R_{0.9}^3 P_s\; .
\end{equation} 

\begin{figure}
\plotone{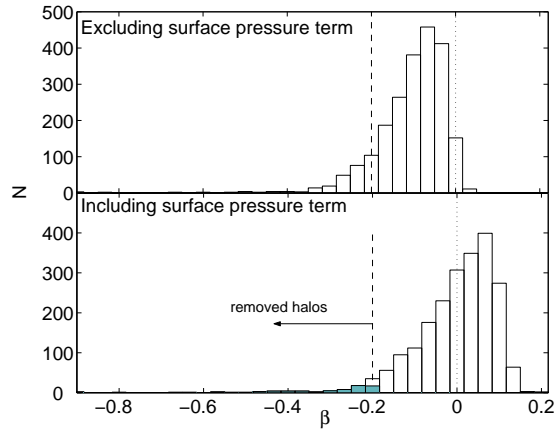}
\caption{Histograms of the distribution of $\beta$ for the halos in
our sample having omitted the 
pressure term $E_s$ from Equation \ref{eq:beta} (top panel). The
distribution peaks at a value just greater
than -0.1, rather than at 0 as would be expected for virialized
halos. Bottom panel--- same as top but having included the pressure
term whilst calculating $\beta$ for each halo. The distribution now
peaks much closer to zero. Also indicated (by the dashed line) is the
value at which we take our lower limit of $\beta$. Halos with $\beta <
-0.2$ (shaded) are removed from our sample for the following analysis.}
\label{fig:beta_hist}
\end{figure}

Figure \ref{fig:beta_hist} compares the virialization distribution
($\beta$) of all the halos in our sample, omitting and including the
surface pressure term respectively. At very low redshift, it is to be
expected that the majority of the dark matter clusters in a simulation
will be dynamically relaxed. By comparing these two distributions it
is clear that our inclusion of the surface pressure term in the virial
theorem is necessary--- the distribution in the lower panel now peaks
very close to zero. The tail of the distribution to negative values of
$\beta$ is largely produced by systems which are infalling and not yet
virialized.

However, the virial equilibrium does not imply that $d^2 I/dt^2 = 0$
instantaneously at all times, but rather that $<d^2 I/dt^2> = 0$,
time-averaged over a period that is long compared to the local
dynamical timescale. Therefore, we expect a roughly symmetric
distribution of $\beta \pm \sigma_{\beta}$ around zero due to those
halos that are oscillating about the virial equilibrium position. It
is difficult to know what exact value of $\sigma_{\beta}$ to expect as
it depends on the state of subdivision of the system.  Thus, as there
are also no halos with $\beta>0.2$, we have picked the value of the
cutoff, $\sigma_{\beta} = 0.2$, so as to leave the remaining
distribution of $\beta$ roughly symmetric about zero, as indicated by
the arrow in the lower panel of Figure \ref{fig:beta_hist}. Although this
cutoff value is essentially determined arbitrarily, we feel that it
sufficiently removes the particularly unvirialized halos in the tail
of the distribution. Applying this criterion to our sample results in
74 (3.4$\%$) of our halos being tagged as `unvirialised'. In Section
\ref{sec:distributions} we demonstrate that these halos have
significantly different physical properties to those that adhere to
our criterium. Reducing the value of $\sigma_{\beta}$ dilutes this
difference, whereas increasing it does not significantly change our
results.

\begin{figure}
\plotone{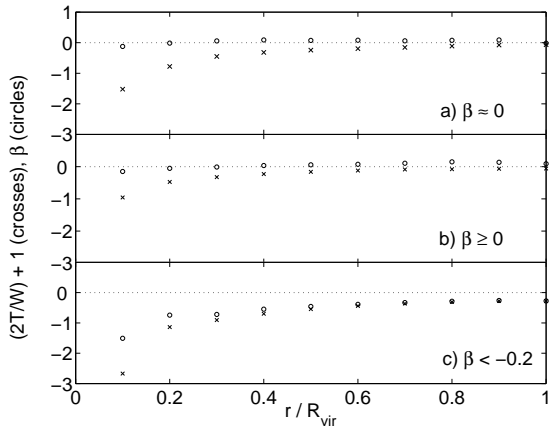}
\caption{Virialization parameter $\beta$ (Eq. \ref{eq:beta}) 
as a function of
radius for three example halos, excluding (crosses) or
including (circles) the surface pressure term $E_s$. }
\label{fig:rvirial}
\end{figure}

\begin{figure}
\plotone{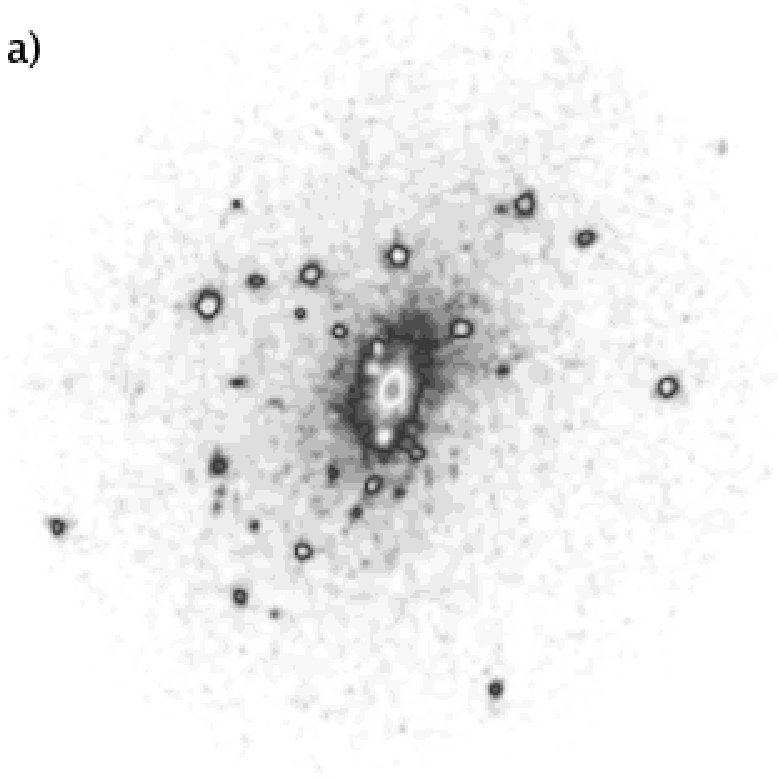}
\plotone{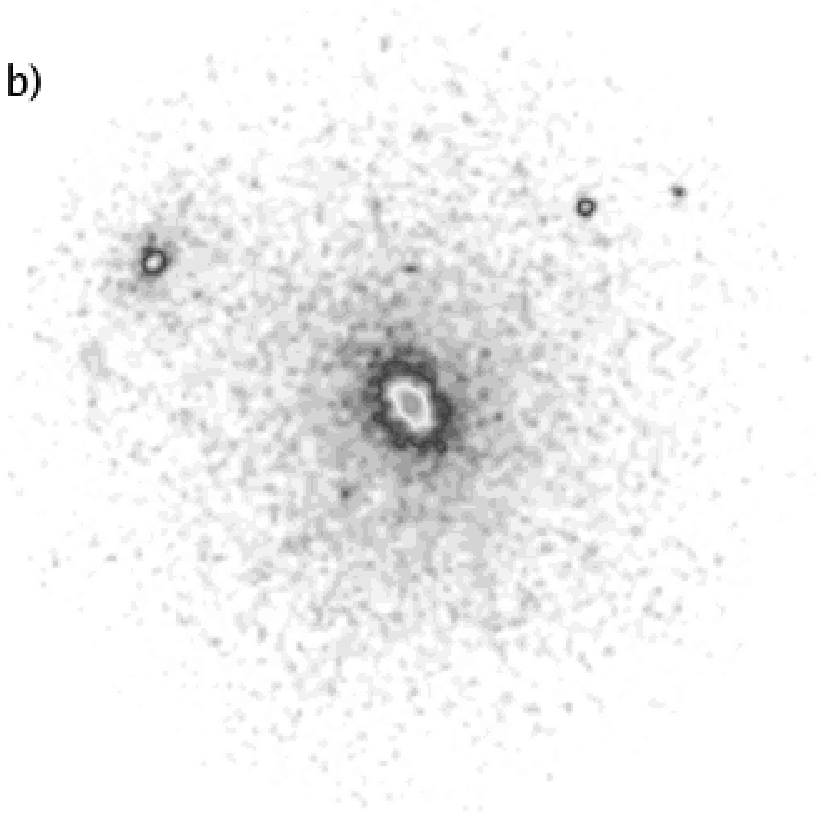}
\plotone{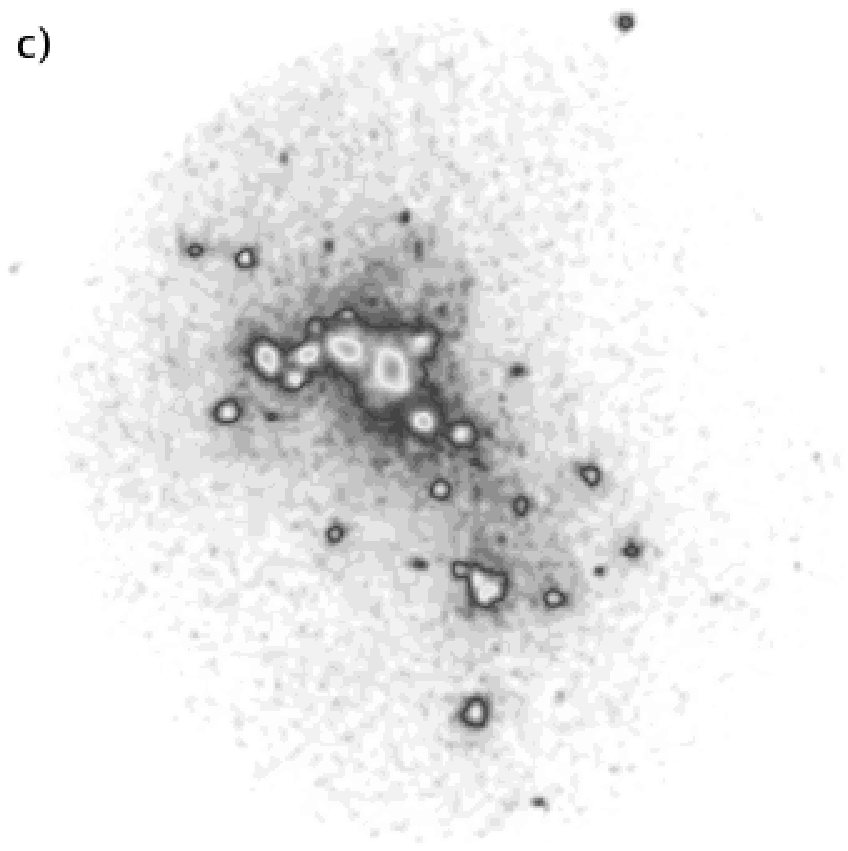}
\caption{Projected density of the three example halos (labeled {\it a
(top), b (middle)} and {\it c (bottom)}) in Figure \ref{fig:rvirial}.}
\label{fig:halopics}
\end{figure}

We should also be able to apply Eq. \ref{eq:beta} at any radii in
order to determine whether the mass within is in dynamically
relaxed. Figure \ref{fig:rvirial} shows the variation of $\beta$ with
radius for three separate halos, including (circles) and omitting
(crosses) the surface pressure correction.  To create this plot, we
have calculated the total kinetic potential energy for all particles
within a certain fraction of the virial radius, $r/R_{\rm vir}$, for
spheres of successively increasing radius $r$ out to $R_{\rm vir}$.
We calculate the surface pressure at each $r$ by applying Equation
\ref{eq:Ps} to the outermost 20$\%$ of particles within that radius.
The three halos have been selected according to their values of
$\beta$ at $R_{\rm vir}$: zero (top), 0.1 (middle) and -0.21 (bottom
panel), using Equation \ref{eq:beta}. Thus the latter would be
identified as being unvirialized, according to our criterium.  Density
plots of the particle distributions for each of these halos can be
viewed in Figures \ref{fig:halopics}.
 
It is clear that the value of 2T/W + 1 (i.e. $\beta$ omitting the
surface pressure) is negative at all radii for all three
halos. However, if we had been able to continue the calculations
beyond the virial radius for the halos in the top two panels in Figure
\ref{fig:rvirial}, it is likely that both would converge to zero. We
note that this is found by \cite{Cole:96} who show in a similar plot
that the ratio 2T/$|W|$ converges to unity for their halos at large
radii (see their Figure 8). By neglecting to include the particles
outside of the virial radius we are omitting an important term in the
virial theorem required to demonstrate that these halos are in fact in
dynamic equilibrium. For these two halos, the surface pressure
corrections demonstrate that the correction achieved through
accounting for the finite pressure and density at each radii result in
values of $\beta$ close to zero throughout.  These halos appear to be
uniformly in dynamical equilibrium, at all radii. The projected
density plots of their particle distributions in Figure
\ref{fig:halopics} (labelled $a$ and $b$) suggest that they are both
relaxed clusters -- a massive central core with smaller orbiting
substructures.

The radial distribution of $\beta$ for the third halo (bottom
panel)--- also the most massive of three--- is considerably more
negative at each radii than the other two halos. This halo has not yet
reached dynamic equilibrium; even with the surface pressure
correction, which is small compared to the halo kinetic energy,
$\beta$ still remains much less than zero at all radii and appears to
converge to a constant value at $r/R_{\rm vir} = 0.7$. Plot $c$ in
Figure \ref{fig:halopics} demonstrates that this halo is clearly a
complicated structure. It appears to have several clumps within its
core, suggesting that it has recently undergone a major merger.

\begin{figure}
\plotone{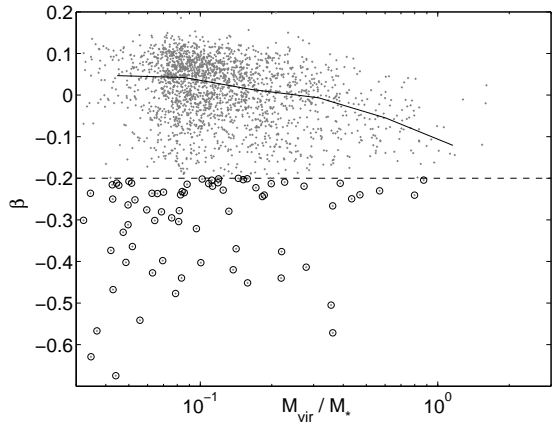}
\caption{Scatter plot of $\beta$ vs the virial mass of each halo
(including the pressure term). The thick
line denotes the median value of $\beta$ in logarithmic mass
bins. Halos with $\beta < -0.2$ (circled) are not used when
calculating the median values in each bin. We take $M_* = 8.0
\times 10^{14} h^{-1} M_{\rm \sun}$.}
\label{fig:massbeta}
\end{figure}

\begin{figure}
\plotone{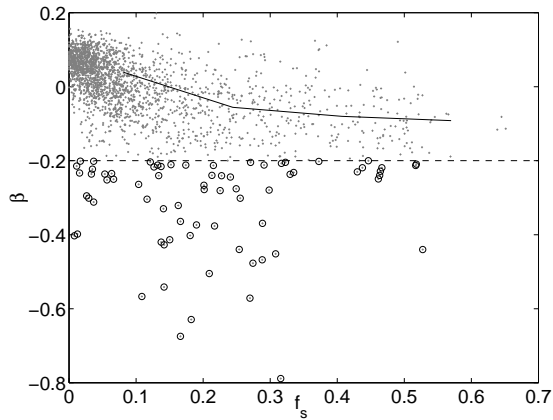}
\caption{Same as Figure \ref{fig:massbeta} but for $f_s$ (the fraction
of mass in substructure) rather than halo mass.
The trend-line 
is calculated having omitted halos with $\beta < -0.2$ (circled).}
\label{fig:fracbeta}
\end{figure}

Figures \ref{fig:massbeta} and \ref{fig:fracbeta} show how $\beta$
varies with halo mass and with the fraction $f_s$ of halo mass in
substructure.  Both of these plots also include lines representing the
median values in mass and $f_{s}$ bins.  Only halos with $\beta >
-0.2$ have been used in calculating the trend line. From Figure
\ref{fig:massbeta}, it appears that there is a weak trend for $\beta$
to decrease (below zero) as halo mass increases.  In accordance with
the current model of hierarchical structure formation
\citep{Lacey:93}, larger and more massive objects have formed more
recently through the merger of smaller structures. Hence, the most
massive dark matter clusters are still in the process of dynamical
relaxation and their constituent particles have an excess of kinetic
energy, which we would expect to dissipate as the cluster settles into
dynamic equilibrium. Halos with larger values of $M_{vir}/M_*$ are,
statisically, younger than those with smaller values of this quantity
and thus are, on average, less `virialised'. The same explanation
applies to Figure \ref{fig:fracbeta}--- which shows $\beta$ decreasing
for halos as the fraction of their mass in substructure increases---
since, as we shall see, the fraction of halo mass in substructure
correlates positively with $M_{vir}/M_*$ and consequently inversely
with halo age.

\begin{figure}
\plotone{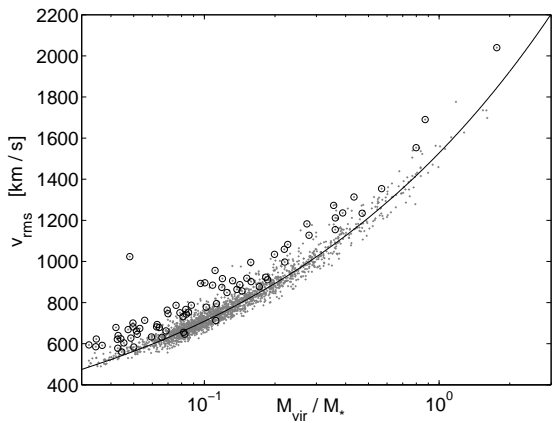}
\caption{Scatter plot of halo mass, $M_{\rm vir}$ against the rms
  particle velocity in each halo. We have circled halos which fail the
  virialization cut (i.e. $\beta < -0.2$). These halos depart from the
  otherwise strong correlation between $v_{rms}$ and $M_{\rm vir}$,
  appearing to have a particularly high kinetic energy per
  particle. The solid line is the NFW relation for mass and $v_{rms}$,
  assuming a halo concentration of 6.5.}
\label{fig:massvrms}
\end{figure}

We will now address the objects that we intend to remove from the
following analysis. These objects have abnormally high kinetic energy
given their mass. They also tend to be lower mass cluster halos
($<10^{14}$ $h^{-1} M_\odot$). In Figure \ref{fig:massvrms} we plot
the particle root-mean-square velocity dispersion against halo
mass. We have ringed all the halos that fail the virialization
cut. As expected, there is clearly a strong correlation between $v_{rms}$ and mass
for the halos we identify as virialized. The solid line represents the
relationship between rms velocity dispersion and mass for an NFW
profile with concentration 6.5 and $\sigma^2_{\rm
\theta}(r)/\sigma^2_{\rm r}(r) = 0.5$, where $\sigma_{\rm \theta}(r)$
and $\sigma_{\rm r}(r)$ are the azimuthal and radial velocity
dispersions respectively (e.g. Eq. 15 of \citet{Lokas:01}). We choose
the above concentration as it is approximately the characteristic
concentration that we obtain for our halo sample (see
Sec. \ref{sec:mass_conc}). Halos that failed the virial criterion
clearly lie above the mass - velocity dispersion relation. As they are
still in the process of relaxing into dynamical equilibrium, possibly
having recently undergone a major merger, it is not surprising that
they have an excess of kinetic energy.

\section{Physical Properties of Halos}

Now that we have defined the halos included in our samples, we can
proceed to discuss the physical properties that we measure. We break
these down into two subsets:

\begin{itemize}
\item{Integral Properties. These are properties that are obtained by
    summing up properties of the individual particles in each halo;
    halo mass, maximum circular velocity (as a measure of halo
    potential) and angular momentum.}

\item{Internal Structure. These describe the distribution of the
    particles within each halo and include concentration, morphology
    and the distribution of substructures.}
\end{itemize}

\subsection{Integral Properties}

\subsubsection{Maximum Circular Velocity}

In order to measure the maximum circular velocity of each halo, we
integrate the NFW density profile (\cite{Navarro:96} to obtain 
\begin{equation}
V^2_{\rm c} = \frac{4\pi G r_i^3 \rho_i}{r}\left[
\ln{\frac{r_i+r}{r_i}} - \frac{r}{r_i+r} \right] \;,
\label{eq:NFWv}
\end{equation}
where $r_i$ and $\rho_i=4\rho(r_i)$ 
are the scale radius and density, and fit to
the circular velocity profile of each halo.  This procedure is
exploited by \citet{Klypin:99} to remove energetically unbound
particles by calculating the escape velocity at the radius of each
particle relative to the halo centre of mass.

The circular velocity profiles for each halo in our halo samples were
measured by radially binning particles into bins of equal particle
numbers. This not only ensures that the Poisson errors associated with
each bin are equal, but also that the radial density of bins is
proportional to the actual density of the halo at any particular
radius. Using the total number of particles within the radius of the
outermost particle in each bin, the circular velocity at that radius
can then be calculated.

We find that the fitted parameters do not vary significantly with the
number of bins used. Furthermore, binning weakens the effect of
substructure on the profile parameters. To quantify the goodness of
fit of the NFW for each halo, we adopt the relative standard deviation
\begin{equation}
\sigma_{rel} = \sqrt{\frac{\sum_{j=1}^n[(V_{\rm c}(r_j) - V_{\rm
    c,nfw}(r_j))/V_{\rm c}(r_j)]^2}{n}} \;,
\label{eq:chi}
\end{equation}
where $V_{\rm c}^2 = GM(<r_j)/r_j$ for bin radius $r_j$, and $V_{\rm
c,nfw}(r_j)$ is the predicted NFW circular velocity at this radius. There has
been much recent discussion (e.g. \cite{Klypin:01}, \cite{Power:03},
\cite{Hayashi:04}) regarding the conditions that must be satisfied in
order to ensure numerical convergence at the innermost radius to which
the circular velocity profiles are computed. Following
\citet{Hayashi:04}, we adopt the following three criteria:
\begin{itemize}
\item{The number of particles enclosed within the innermost radius,
$r_{\rm in}$, is sufficient to ensure that five times the local
collisional relaxation timescale is longer than the age of the
universe, which is of the order of the orbital timescale at the virial
radius \citep[see for example][]{Binney:87}
\begin{displaymath}
\frac{t_{relax}(r_{\rm in})}{t_{circ}(R_{\rm vir})} =
\frac{N(r_{\rm in})}{8ln(N(r_{\rm in}))}\frac{r_{\rm in}/V_c(r_{\rm in})}{r_{200}/V_{200}}
\end{displaymath}
\begin{equation}
= \frac{\sqrt{200}}{8}\frac{N(r_{\rm in})}{ln(N(r_{\rm in}))}\left[\frac{\bar{\rho}
      (r_{\rm in})}{\rho_{\rm crit}}\right]^{-1/2}
\geq 5\; ,
\label{eq:converge}
\end{equation} 
where $N(r_{in})$ is the number of particles and $\bar{\rho}(r_{in})$
the mean density within the radius $r_{in}$. It takes at least 15
relaxation times for core collapse to occur and 200 for the system to
evaporate, hence we have chosen 5 as a convergence criterion to allow
us to probe the inner regions of each halo, without being affected by
numerical instabilities.}
\item{The innermost radius is greater than 4 softening lengths
    ($\epsilon = 3.2 h^{-1} kpc$ in this simulation)}
\item{There at least 100 particles within the innermost radius}
\end{itemize}
As in \citet{Hayashi:04}, we find that the first constraint is the
strictest for the vast majority of our halos, and virtually every halo
that satisfies the third constraint automatically satisfies the second
as well.

\subsubsection{Spin}
The angular momentum of simulated dark matter halos is typically been
parameterized by the dimensionless spin parameter 
\citep{Peebles:69,Peebles:80}
\begin{equation}
\bar\lambda \equiv \frac{J\sqrt{E}}{GM^{5/2}}\; ,
\end{equation}
where $M$ and $E$ are the total mass and energy of the system, $G$ is
the gravitational constant and $J$ is the total angular momentum of
the halo. This definition of spin is directly related to the ratio of
the centrifugal to gravitational acceleration at the equator of a
homogeneous rotating body, or the amount of angular momentum required
for rotational support. 

\citet{Peebles:69} showed that the amount by which a uniformly
rotating polytrope is deformed, or how oblate it is, is related to
$\bar\lambda$ in the following way:
\begin{equation}
\frac{a - c}{a} \approx
\frac{J^{2}|E|}{G^{2}M^{5}} = \bar{\lambda}^{2} \; ,
\label{eq:polytrope}
\end{equation}
i.e. the flattening of a rotating sphere is directly proportional to
the square of its associated spin parameter, where $a$ is the major
and $c$ the minor principle axis of the halo.

As we outlined in Section 1, the total energy component of
$\bar\lambda$ is difficult to compute due to the necessity of
accounting for the surface pressure at the outer radius of each
halo. This is especially the case for halos in high density regions of
the simulation. Therefore, \citet{Bullock:01} introduced an
alternative parameterization,
\begin{equation}
\lambda = \frac{J}{\sqrt{2}MVR}\; ,
\end{equation}
which sidesteps this difficulty by removing the energy dependence,
thus making it more practical to calculate from simulated data. We
adopt this version of the spin parameter, using the virial mass and
radius for $M$ and $R$ respectively and calculating the circular velocity
component, $V^2 = GM(R_{\rm vir})/R_{\rm vir}$. This retains some
consistency between the two approaches, as $\lambda$ reduces to
$\bar\lambda$ when measured at the virial radius of a truncated
isothermal sphere.  \citet{Hetznecker:05} have performed a
comprehensive study of the differences between the two definitions of
halo spin. In agreement with \citet{Bullock:01} they find there to be
little difference in the overall distributions of both $\bar\lambda$
and $\lambda$ at z=0. However, they also demonstrate that whilst
$\lambda$ ($\lambda^{\prime}$ in their notation) does not vary
noticeably with redshift, $\bar\lambda$ increases significantly as
redshift decreases. This discrepancy appears to be due to the impact
of the continuous acquisition of mass through accretion, which
increases $\bar\lambda$ but slowly decreases $\lambda$.

\subsection{Internal Structure Properties}

\subsubsection{Substructure}
The procedure we employ to identify substructure and build the
associated family trees for each cluster halo is briefly described in
Section \ref{sec:defsam}. The full procedure is completely described
in \citet{Weller:04}. One important feature of this algorithm is that
near neighbor subhalos, which are gravitationally bound to one
another, are associated together. If this is not done, the subhalo
distribution might depend on the geometrical smoothing lengths used in
the definitions of subhalos. Here we will present the subhalo
mass-function for the 2159 cluster halos in our sample, including
those tagged as `unviriliased' .

\subsubsection{Concentration}

Halo concentration is normally defined as the ratio of some measure of
the outer radius of a halo to an inner `scale' radius, normally chosen
to be that of the NFW profile, i.e. when the logarithmic slope of the
density profile = -2. Previous studies have used various definitions
of the outer radius of a halo: normally it is chosen to be the radius
at which the halo density becomes some factor of the critical or mean
background density of the universe at the appropriate redshift. In
this study we use the virial radius to determine halo concentration,
so $c \equiv R_{\rm vir}/r_i$. When comparing our results with studies
that have adopted an alternative definition of halo concentration
(e.g. relative to the mean background, rather than the critical density)
we use a conversion factor to correct for the difference.

\subsubsection{Morphology}
In order to measure halo morphology, we calculate the moment of
inertia tensor for each halo, $I_{ij} = m_{p}\sum_{k} r^{i}_kr^{j}_k$,
summed over all particles in the halo, where $m_{p}$ is the particle
mass in the simulation and $r^{i}_k$ is the x, y or z component of the
distance of the k-th particles from the cluster center of mass. We
then find the magnitude and orientation of the principle axes for each
halo by diagonalising $I_{ij}$ and calculating the associated
eigenvalues and eigenvectors respectively.  We define the axis ratios
as in \citet{Warren:92a}, denoting $a$ as the longest axis, $b$ the
intermediate axis, and $c$ the shortest axis of the moment of inertia
ellipsoid.

Previous studies have calculated the inertia tensor using an iterative
scheme in which only particles in a particular sphere or shell are
used to calculate $I_{ij}$. This is then repeated, using the ellipsoid
or shell determined by the principle axis vectors from the
diagonalized inertia tensor, until until the procedure converges
\citep{Warren:92a}. In agreement with \citet{Jing:02} and
\citet{Bailin:04} we find that this approach often does not converge
in high resolution simulations where the over-merging problem--- which
tends to erase substructure in low resolution simulations--- does not
apply. \citet{Jing:02} adopt a novel approach in which they fit halos
shapes to isodensity contours. However, \citet{Allgood:05} find that
this has the effect of suppressing the impact of large substructures
near the halo core, making higher mass halos appear more prolate than
they really are. 

\section{Distribution of Halo Properties}\label{sec:distributions}

In this Section we describe the distributions of the halo properties
listed above for the halos in our sample. Unless specifically stated
otherwise, we plot separately the properties of the halos tagged as
`virialized' and, where relevant, those tagged as `unvirialized'.


\subsection{Spin}

As has been established in earlier studies \citep{Barnes:87,
Warren:92a, VandenBosch:98, Bullock:01, VandenBosch:02, Gardner:01,
Vitvitska:02, Peirani:04}, the distribution of $\lambda$ is well fit
by a log-normal distribution
\begin{equation}
P(\lambda) = \frac{1}{\lambda \sqrt{2\pi} \sigma} \exp\left(-\frac
{\ln^{2}(\lambda/\lambda_{0})}{2\sigma^{2}}\right)\; .
\label{eq:lognorm}
\end{equation}

\begin{figure}
\plotone{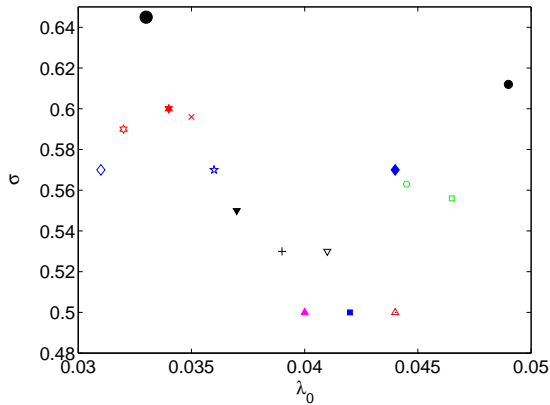}
\caption{Values of $\lambda_0$ and $\sigma$ as in the literature (see
text). The large black solid circle is the result of the analysis
presented here for our main halo sample, the smaller black solid
circle for our `unvirialised' sample.}
\label{fig:lamsig}
\end{figure}

In their analysis, \citet{Bullock:01} find ($\lambda_0$,$\sigma$) =
(0.042,0.5) (blue, solid square in Figure \ref{fig:lamsig}), whilst
\citet{Gardner:01} find (0.044,0.5) (red, upward-pointing triangle)
for the halos that have undergone major mergers during their
formation, and (0.035,0.596) (red cross) for those that have grown
through accretion. In a similar study, \citet{Peirani:04} found
(0.036,0.57) (blue, five-pointed star) over all the halos in their
sample, whilst $\lambda_0$ $\approx$ 0.044 (filled blue diamond) for
the merger halos and $\lambda_0$ $\approx$ 0.031 (blue diamond) for
the accretion dominated halos (see their Figure 7).  Likewise,
\citet{Hetznecker:05} obtain (0.039,0.53) for their entire halo sample
(black plus sign), (0.041,0.53) for their merger halos (black,
downward pointing triangle) and (0.037,0.55) for the accretion
dominated halos (filled, black downward pointing
triangle). \citet{Vitvitska:02} find (0.0445,0.563) from their
numerical simulations (green circle) and (0.0465,0.556) (green square)
from their random walk model (based on the simulations). Whilst
studying the dependence of $\lambda$ on environmental density,
\citet{Avila-Reese:05} obtain (0.032, 0.59) for their `Void' halo
sample (open, red six-pointed star) and (0.034,0.60) for their `Field'
halo sample (filled, red six-pointed star). Finally, by modeling the
acquisition of spin by halos from both the orbital angular momentum of
merging satellites and through the effects of a large-scale tidal
field \citep[tidal torque theory--- see for
  example][]{Peebles:69,White:84}, \citet{Maller:02} find that both
scenarios produce a log-normal distribution of halo spin with
$\approx$ (0.04,0.5) (filled magenta triangle). Overall, most recent
studies fix $\lambda_0$ in the range 0.031-0.045, with dispersions
varying between 0.5-0.6.

\begin{figure}
\plotone{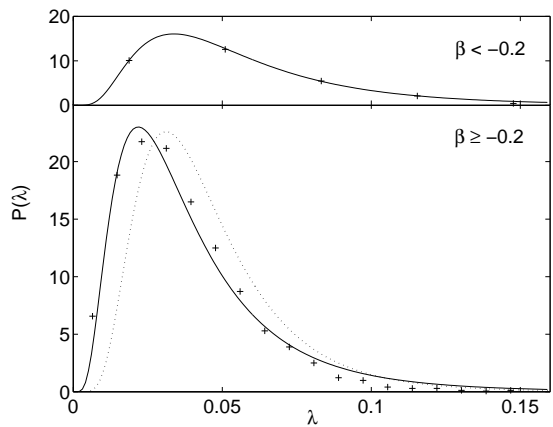}
\caption{({\it bottom panel}) Distribution of the halo spin parameter,
  $\lambda$ (crosses) and fit of Equation \ref{eq:lognorm} (solid
  line), where $\lambda_0$ = 0.033 and $\sigma$ = 0.645. For
  comparison, we plot a distribution with $\lambda_0 = 0.04$ and
  $\sigma=0.5$ \citep{Maller:02}. ({\it top panel}) Same, but for
  those tagged as `unvirialized', where $\lambda_0$ = 0.049 and
  $\sigma$ = 0.612.}
\label{fig:hist_lam}
\end{figure}

The value we obtain for $\lambda_0$ (0.033--- see lower panel in
Figure \ref{fig:hist_lam} and the large filled black solid circle in
Figure \ref{fig:lamsig}) is in broad agreement with these studies,
however we obtain a slightly broader distribution with $\sigma$ =
0.645.  For comparison, we also plot the distribution obtained by
\citet{Maller:02} (dotted line in Figure
\ref{fig:hist_lam}). Interestingly, our results appear to be very
similar to those obtained by \citet{Gardner:01}, \citet{Vitvitska:02}
and \citet{Peirani:04} for the halos in their sample that have grown
through accretion only, having not undergone any major mergers during
their formation.  \citet{Vitvitska:02} and \citet{Gardner:01} both
suggest that increased merging activity will result in a higher halo
angular momentum--- a result that is verified by \citet{Peirani:04},
who monitor angular momentum as halos form in their
simulations. However, the higher dispersion that we measure than any
previous study does indicate that there are still a significant number
of high $\lambda_0$ halos in our sample, indeed, over 300 halos have
$\lambda_0$ $>$ 0.06.

The top panel in Figure \ref{fig:hist_lam} describes the spin
distribution for halos that we have tagged as `unvirialized', using
the criterion outlined in Sec. \ref{sec:virial}. This is also well
described by Eq. \ref{eq:lognorm}, with $\lambda_0$ = 0.049 and
$\sigma$ = 0.612 (small, black filled circle in Figure
\ref{fig:lamsig}). A larger value of $\lambda_0$ is consistent with
the idea that these halos have not yet reached a state of dynamical
equilibrium and have greater kinetic energy (partly in the form of
angular momentum) than their virialized counterparts.

\subsection{Mass function and fraction of mass in substructure}
\label{sec:subhist}

\begin{figure}
\plotone{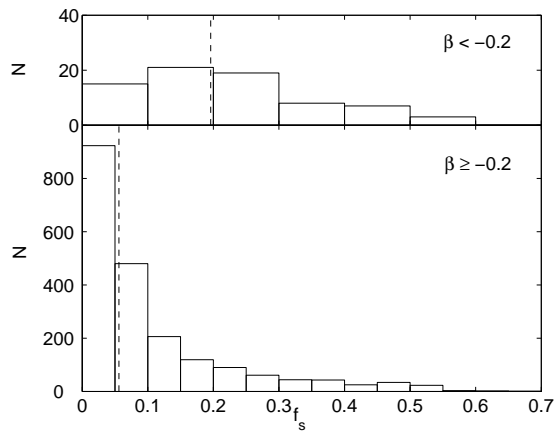}
\caption{({\it bottom}) Distribution of the fraction of mass in
substructure for our main halo sample. ({\it top}) The equivalent
distribution for those halos that were tagged as 'unvirialised'}
\label{fig:frac_hist}
\end{figure}

The bottom panel in Figure \ref{fig:frac_hist} displays the
distribution of the fraction of mass contained in substructure for our
main halo sample. The median value of $f_s$ is 0.056, where $f_s$
ranges between $0 \leq f_s \leq 0.65$. Note also that there is a large
tail to high values of $f_s$. The upper panel of Figure
\ref{fig:frac_hist} demonstrates that the unvirialised halos do have
on average more substructure, with a median value of $f_s =
0.196$. This indicates that these are likely to be young clusters in
which there has not been sufficient time for the halo core to remove a
significant fraction of subhalo mass through tidal ablation. Both
\cite{Gill:04b} and \cite{DeLucia:04}, who resimulate 8 and 11 cluster
halos respectively, and \cite{Gao:04}, who resimulate halos over a
wider range of masses, obtain mean values for $f_s$ close to the ones
presented here.

\begin{figure}
\plotone{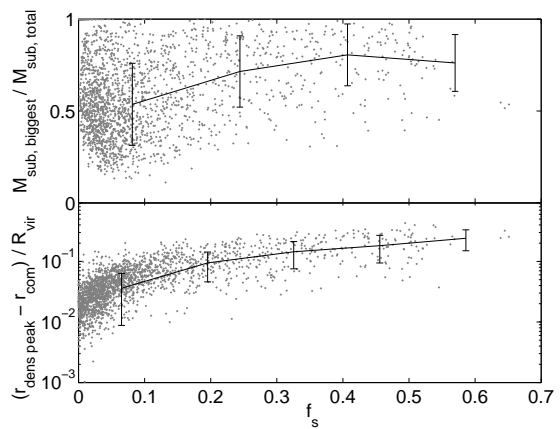}
\caption{({\it top}) 
Ratio of the mass in the most massive subhalo 
to the total  mass in substructure,
as a function of the total
fraction of mass in substructure, $f_s$. The plot demonstrates that for
halos with $f_s > ~0.3$ the substructure is largely dominated by a
single massive subhalo of mass comparable to that of the mother
halo. ({\it bottom}) Scatter plot of the distance between the point of
peak density and the center of mass, as a fraction of the virial
radius, for each halo against $f_s$.}
\label{fig:massivesub}
\end{figure}

In the distribution in the lower panel of Figure \ref{fig:frac_hist}
there is tail of high $f_s$ halos. These are halos that have very
recently undergone a major merger (although they have passed our
virialisation criterion) and have a dual core at their center--- the
mother halo and a slightly less massive subhalo.  In the upper panel
of Figure \ref{fig:massivesub} we plot the mass of the largest subhalo
as a fraction of the total mass in substructure against $f_s$.  It is
clear that, on average, halos with a large fraction of mass in
substructure host a subhalo of almost equivalent mass to the
mother. This is also demonstrated in the lower panel, where we plot
the distance between the center of mass of the halo and its position
of peak density, as a fraction of the virial radius, against
$f_s$. The density peak, by definition, resides in the mother halo.
However, for halos with an especially massive subhalo, the center of
mass will lie somewhere in between the density peak of the mother and
that of the subhalo. For these halos, an NFW profile provides a poor
fit (see Figure \ref{fig:NFW_fit}) to the radial density profile due
to the impact of the large subhalo.  The correlation indicates that
the center of mass moves further from the peak density position as
$f_s$ increases--- suggesting that much of the substructure is
clustered on one side of the halo. This can be explained if halos with
large $f_s$ are more likely to have one large piece of substructure,
rather than lots of smaller subhalos that just happen to be located on
one side of the mother halo.

\begin{figure}
\plotone{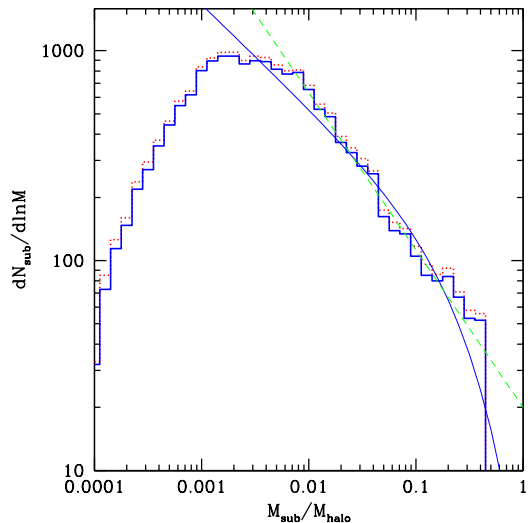}
\caption{Mass distribution of subhalos from the 2159 most massive
halos in our simulation as a function of the ratio of subhalo mass to
cluster mass. The dotted histogram is the distribution of the complete
sample, the solid histogram for the 2085 virialised halos only. The
solid line is the Schechter function fits as discussed in the
text. The dashed line is obtained for just fitting the slope. In both
distributions halos with less then 30 particles have been dissolved
into their mothers, thus setting a lower mass limit for the subhalos
in our sample.}
\label{fig:massfunct}
\end{figure}

In Figure \ref{fig:massfunct} we plot the mass distribution of the
subhalos we obtain from our sample of 2159 clusters, including those
tagged as `unvirialised'.  The dotted histogram is the distribution
for all the sample halos, while the solid histogram only includes the 
2085 virialized halos. The number density $N_{\rm sub}$ of subhalos of mass
$M_{\rm sub}$ can be described by a Schechter function
\citep{Press:74a,Schechter:76a}: \beq \frac{dN_{\rm sub}}{d\ln(M_{\rm
sub})} = N_* \left(\frac{M_{\rm sub}}{M_{*sub}}\right)^{-\alpha}
\exp\left[-\frac{M_{\rm sub}}{M_{*sub}}\right]\, ,
\label{eqn:PS}
\eeq 
which is a power law with an exponential cut-off at large mass
scales, where $N_*$ is a normalization constant.  In
Figure~\ref{fig:massfunct} we show the mass function of subhalos
(solid line) where the substructure mass is expressed in units of the
halo mass; this brings potentially different mass halos onto an
equal footing. At a mass scale of about $M_{\rm sub} = 3\times10^{-3}
M_h$ the function drops, because at this stage our finite sample size
and the lower mass limit of our identified halos of $7.6\times 10^{10}
h^{-1} M_{\rm \sun}$ truncate the distribution.

We fit this distribution up to the truncation point with the Schechter
function (Equation~\ref{eqn:PS}), varying all three parameters, and
obtain $\alpha = 0.5$ and $M_{*sub} = 0.3 M_h$ (solid line).  However,
there are large correlations or parameter degeneracies amongst
$\alpha$, $N_*$, and $M_{*sub}$, which lead to marginalized errorbars
in $\alpha$ of $\pm 0.2$ and in $M_{*sub}$ of $\pm 0.7 M_h$. In order
to compare with previous work we omit the cut-off, $M_{*sub}$, and
just fit for the slope and amplitude. In this case we obtain $\alpha =
0.75$ (dashed line).  Still there is a remaining correlation between
$\alpha$ and $N_*$ and the marginalized errorbar on the slope is $\pm
0.12$. If we just fit for the slope we obtain $\alpha = 0.75 \pm
0.12$.

Furthermore, we find the distribution becomes flat below $M_{\rm sub}
\approx 3\times10^{-3} M_h$. Others
\citep{Ghigna:00a,Springel:01a,DeLucia:04, Gao:04} have found similar
results with regard to the power law component; but we have added an
exponential cut-off at the high mass end, which is only possible of
our large halo sample.  We note the cutoff in high mass objects away
from the simple power law, as is predicted from the Schechter
prescription \citep{Schechter:76a}. Such a cutoff is of course
essential, as there cannot, by definition, exist subhalos more massive
than the mother halo. At this stage the distribution of subhalos is
steeper than the global distribution of all the halos in our
simulation \citep{Weller:04} by almost $0.2$ in the exponent. Our
explanation for this is that while the original input distribution to
a growing halo must correspond to the global average distribution, the
more massive subhalos are more quickly destroyed as dynamical friction
subjects them to more rapid tidal ablation, thus steepening the
profile.
 
Note that the slopes of the Schechter function obtained here are lower
than the ones obtained in \citet{DeLucia:04}, which are in the range
$\alpha\approx 0.85 - 1.13$. However if we fit for the slope alone the
2-$\sigma$ region of our fit is $0.51 < \alpha < 0.99$, which overlaps
with this range of \citet{DeLucia:04}. The difference between our
result and previous discussions \citep{Ghigna:98a,Moore:99a,Klypin:99,
Helmi:02,DeLucia:04} is due to the fact that we include {\em all}
halos, regardless of their morphology, while previous studies picked
halos which are not in merger states. While the choice of relaxed
halos clearly produces a bias, our sample might be biased as well,
since one expects more mergers among the most massive halos in the
simulation than in a randomly selected sample.

\subsection{Concentration}

It has been demonstrated by \citet{Jing:00} and \citet{Bullock:01b} that
the distribution of halo concentrations at any particular redshift can
be well described by a log-normal distribution, of dispersion $\sigma
\approx 0.22$. \citet{Dolag:04} showed that this result is
independent of the adopted cosmological model. It is also thought that
the dispersion in concentrations is related to the dispersion of halo
formation epochs \citep{Wechsler:02}. 

\begin{figure}
  \plotone{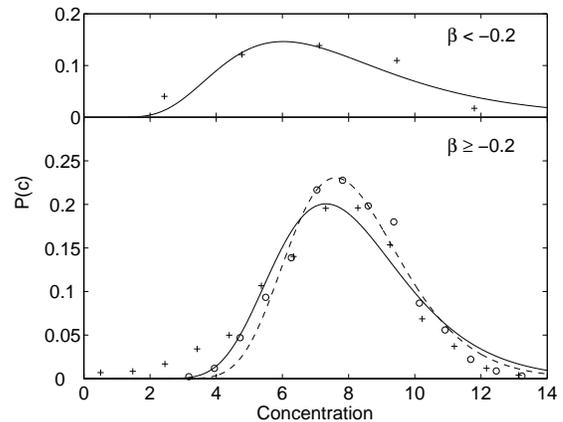}
\caption{({\it bottom}) Distribution of halo concentrations, including
(crosses) and omitting (circles) halos with $f_s \geq 0.3$. Solid and
dashed lines are the best fit log-normal distributions. ({\it top})
The distribution for halos tagged as `unvirialised' according to our
virialisation criterion (Sec. \ref{sec:virial}).}
\label{fig:conc_hist}
\end{figure}

\begin{figure}
\plotone{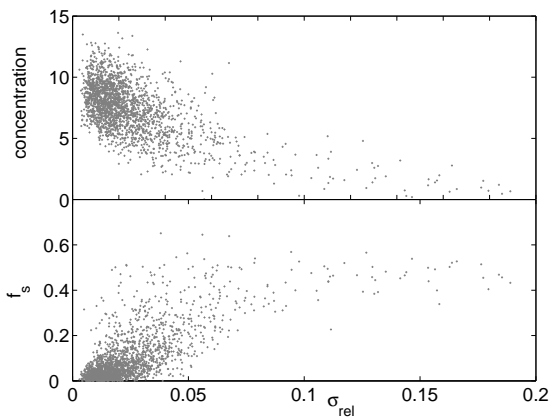}
\caption{Scatter plots of halo concentration (upper panel) and the
fraction of mass in substructure (lower) with respect to the goodness
of the NFW profile fit (as defined by Equation \ref{eq:chi}).}
\label{fig:NFW_fit}
\end{figure}

Figure \ref{fig:conc_hist} shows the best fit log-normal distribution
(solid line) to our halo sample.  At first glance, this Figure appears
to suggest that our halo sample (crosses) does not conform to such a
distribution. This is mainly due to the excess of halos with a very
low concentration ($<3$). These halos appear to have significantly
more mass in their outer regions than is suggested by an NFW
profile. Indeed, as demonstrated in the upper panel of Figure
\ref{fig:NFW_fit}, the NFW profile provides a very poor fit to these
halos; the reason for this is that these halos have a considerable
fraction of their mass in substructure (Figure \ref{fig:NFW_fit},
lower panel).  Halos with a concentration of less than $\sim 2.1$ are
those that have a massive (bound) daughter at or near its virial
radius. For these halos, the circular velocity profile continues to
rise beyond $R_{\rm vir}$. The open circles in Figure
\ref{fig:conc_hist} represent the distribution of halo concentrations
once all halos with $f_s > 0.3$ are removed.  It is clear that this
sub-sample is well described by a log-normal distribution (dashed
line) with a dispersion $\sigma_c = 0.22$.

The top panel displays the distribution of concentrations for the halos
tagged as `unvirialized'.  The distribution is also well described by
a lognormal distribution, but peaks at a far lower concentration than
the main halo sample. In these halos, much of the matter that will
eventually be accreted onto the halo core is still in the outer
regions, thus reducing the concentration. Furthermore, they also have
(on average) a higher fraction of their mass contained within
substructures, many of which are located outside the central regions
of the halo.


\subsection{Morphology}
\label{sec:morphology}

\begin{figure}
\plotone{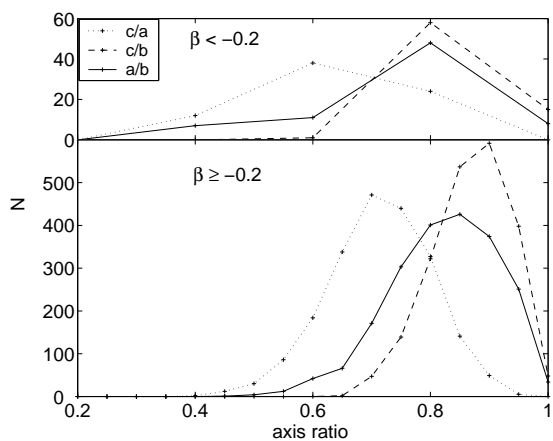}
\caption{({\it Bottom:}) histogram of the distributions of the halo axial
ratios, b/a (solid line), c/b (dashed) and c/a (dotted). The median
values agree quite well with previous studies. ({\it Top:}) same,
but for the halos tagged as `unvirialised'.}
\label{fig:axisratios}
\end{figure}

Figure \ref{fig:axisratios} displays the distributions for each
principle axis ratio.  The ratio of the intermediate to minor axes is
in general greater than that of the major to intermediate axes. This
implies that the halos are more prolate than oblate --- a result found
by several previous studies \citep{Davis:85, Warren:92a, Cole:96,
Kasun:04, Bailin:04}. None of the three distributions resemble a
normal distribution, unlike those found by \citet{Jing:02}, who
resimulate a small number of halos over a broad mass range
($10^{12}-10^{14} M_{\rm \sun}$).  As both \citet{Bailin:04} and
\citet{Kasun:04} have found, we see a tail towards very low axis
ratios, corresponding to extremely flattened halos.

The mean values we obtain for the axis ratios are b/a = 0.817 $\pm$
0.098, c/b=0.867 $\pm$ 0.067 and c/a = 0.707 $\pm$ 0.095 respectively.
Again, these values agree quite well with the axis ratios found by
\citet{Warren:92a, Cole:96} and \citet{Kasun:04}.  In the case of
\citet{Warren:92a} we find our results match those they obtain for the
largest radius used in their iterative ellipsoidal fitting
technique.  \citet{Dubinsky:91} obtain a very low mean c/a ratio of
0.5 for the isolated collapse of halos with 100kpc radius. However,
both \citet{Dubinsky:91} and \citet{Warren:92a} also find that the
halos become steadily more spherical as the length of the semi-major
axis of the shell was increased in their analysis.  From the Gaussian
fit to their distribution of c/a, \citet{Jing:02} find a mean value of
0.54 with a dispersion of 0.113. They then construct a conditional
probability distribution for b/a over certain ranges of c/a, 
yielding
a maximum likelihood value of 0.77 for b/a, given their
mean value of c/a. \citet{Bailin:04} break each of their halos down
into concentric spheres before measuring the inertia tensor in each
shell (rather than for the entire halo) so it is difficult to compare
directly the values we obtain for the mean axis ratios.  Comparing to
their most massive halos ($10^{13} - 3\times 10^{14} M_{\rm \sun}$) we find
that the axis ratios we obtain are greater (i.e. closer to 1), 
with the exception of those they obtain using only
the innermost shells (see their Figure
3). Finally, recent results by \citet{Allgood:05} measure a c/a ratios
of $0.6 \pm 0.1$ for their galaxy mass halos. These results are also
in reasonable agreement with observational work by
\citet{Basilakos:00}, who measure the projected cluster ellipticities
of APM galaxy clusters based on moments of the smoothed galaxy
distribution. They find that a prolate spheroidal model fits the data
best and (without excluding clusters that show significant
substructure) measure a c/a axis ratio of 0.65 $\pm$ 0.15.

\begin{figure}
\plotone{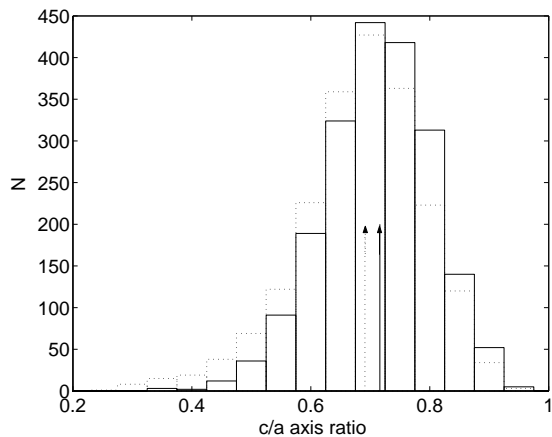}
\caption{Solid histogram: distribution of the c/a axis ratio.  Dotted
histogram: the same, but substructure just outside of the virial
radius is included in the calculation of the inertia tensor for each
halo. The solid and dotted arrows indicate the median values of each
distribution. There is a clear shift towards lower values of c/a when
including outlying substructure, suggesting a slightly less spherical
morphology.}
\label{fig:allsubs}
\end{figure}

Overall, the halos in our simulations are slightly more spherical than
those in previous studies. This is possibly due to our selection
policy of removing from our analysis subhalos that have their center
of mass just outside of the virial radius of the halo.  Large
pieces of substructure located just outside of the halo would,
if included as part of the halo, have the effect of making it appear
more prolate and thus decreasing the c/a axis ratio.  In order to test
this we have repeated the analysis, but now including all substructure
that has at least one particle inside of the virial radius of the
halo. Figure \ref{fig:allsubs} compares the c/a distributions
obtained from this new dataset with the original. It is clear that the
new histogram has many more halos with a low c/a axis ratio than the
original, especially in the tail. The new mean values for the axis
ratios are b/a = 0.79 $\pm$ 0.121, c/b = 0.859 $\pm$ 0.071 and c/a =
0.678 $\pm$ 0.113. The fact that the mean minor-to-intermediate axis
ratio has decreased the least suggests that the halos have become more
prolate with the added substructure.

The top panel in Figure \ref{fig:axisratios} shows the distribution of
the axis ratios for the halos that we have tagged as being
`unvirialized' according to our criterion outlined in
Sec. \ref{sec:virial}. They indicate that these halos are less
spherical than their more relaxed counterparts, probably due to their
larger fraction of mass contained in substructure. We investigate
further the impact of substructure on halo morphology in Section
\ref{sec:impactsub}.

\begin{figure}
\plotone{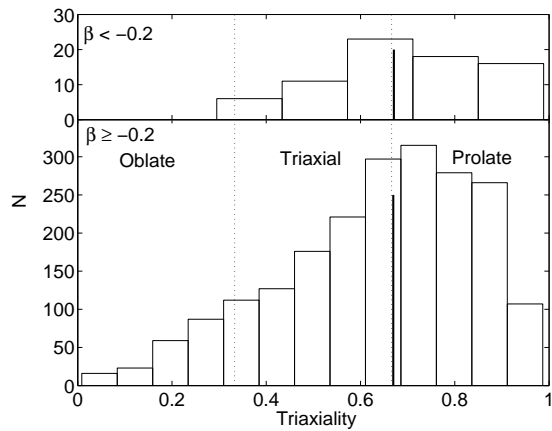}
\caption{Distribution of the triaxiality, T (see Equation \ref{eq:T}),
for our main halo sample ({\it bottom}), and those tagged as
'unvirialised' ({\it top}). Thick black lines indicate the medians of
each distribution.  Both panels show that there is a clear preference
for halos to adopt a more prolate, rather than oblate, morphology.}
\label{fig:hist_triax}
\end{figure}

Another way in which the shape of simulated dark matter halos can be
classified is in terms of their triaxiality, T \citep{Warren:92a,
  Bailin:04}, defined as:
\begin{equation}
T \equiv \frac{a^{2} - b^{2}}{a^2 - c^2} \;.
\label{eq:T}
\end{equation}
The full range of triaxiality falls into three regimes: oblate (or
`pancake-like') halos correspond to T $<$ 1/3, prolate (or
`cigar-like) to T $>$ 2/3 and fully triaxial halos to the intermediate
range. From Figure \ref{fig:hist_triax}, it is clear that the majority
(52$\%$) of halos lie in the T$>$2/3 range and are therefore more
prolate than oblate, with the distribution peaking at T$\sim$0.75. A
significant number of halos (38$\%$) are completely triaxial --- the
principle axis lengths are distributed over a significant range. 

\section{Correlations with Mass}
\label{sec:corrmass}

In this Section we investigate how the halo properties we measure
correlate with halo mass. This is important for determining whether
certain halo characteristics are self-similar, or whether they vary
significantly depending on the age or phase of evolution of a
halo. All the results presented in this Section are for the
`virialised' halos in our sample only. As most previous studies have
found \citep{Davis:85, Warren:92a, Avila-Reese:05}, we see no
measurable correlation between halo mass and spin.

\subsection{Maximum Circular Velocity}
\label{sec:mass_vmax}

\begin{figure}
\plotone{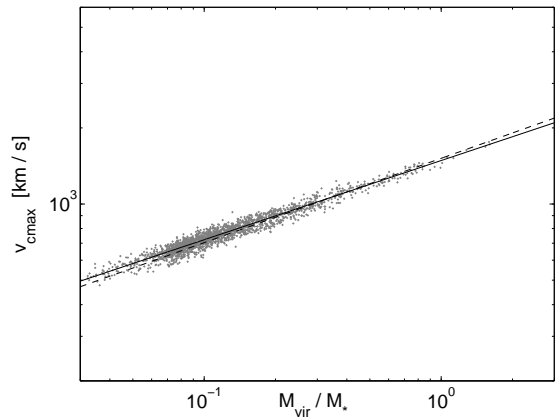}
\caption{Maximum circular velocity as a function of halo mass, having
  removed halos for which the NFW profile provides a poor fit due to
  large amounts of substructure. Straight line is best fit of function
  $V_{\rm cmax} = A(M_{\rm vir}/M_*)^\alpha$, with $\alpha = 0.31\pm
  0.08$. The dashed line is the NFW {\it prediction} based on our
  measured correlation between halo mass and concentration (Equation
  \ref{eq:c}).}
\label{fig:mass_vmax}
\end{figure}

Figure \ref{fig:mass_vmax} plots maximum circular velocity as a
function of virial mass. We have removed all halos with $f_s > 0.3$ as
the NFW profile provides a poor fit to many of these halos due to
their large quantity of substructure (see Figure
\ref{fig:NFW_fit}). Fitting a simple power-law $V_{cmax} \propto
M^{1/\alpha}$, we measure $\alpha = 0.31\pm 0.08$ (solid line). This
is very close to the relationship obtained by \citet{Kravtsov:04},
\citet{Bullock:01b}, and \citet{Hayashi:03} (for their subhalos in the
latter). \citet{Bullock:01b} explain the departure from that predicted
by standard scaling of the virial parameters, $M_{\rm vir} \propto
V_{\rm vir}^{3}$, through the observed mass dependency of the
concentration parameter--- a result that we also find (Section
\ref{sec:mass_conc}).  For the NFW profile, $V_{\rm cmax} \propto
M_{vir}^{1/3}f(c)$, where $f(c)$ is a function of halo concentration
which in turn depends on $M_{vir}$ (see, for example,
\citet{Avila-Reese:99}). In Section \ref{sec:mass_conc} we find a
power-law relation between $c$ and $M_{vir}$. The dashed line in
Figure \ref{fig:mass_vmax} represents the corresponding relationship
between $M_{vir}$ and $V_{\rm cmax}$.  

\citet{Kravtsov:04} follow the evolution of dwarf halos as they first
accrete mass in isolation and then undergo significant mass loss
through tidal stripping whilst being captured by more massive
neighboring halos. They find the M-$V_{\rm cmax}$ relationship holds
throughout the differing phases of subhalo evolution. Hence, it
appears that this correlation is independent of halo mass and
formation epoch.  This helps to explain the tightness of the relation
in Figure \ref{fig:mass_vmax}.

\begin{figure}
\plotone{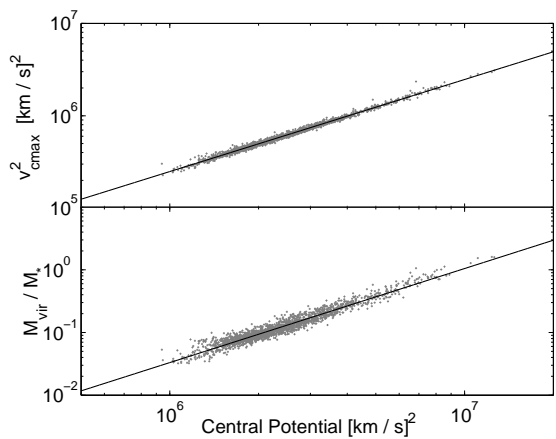}
\caption{Maximum circular velocity ({\it top}) and halo mass ({\it
bottom}) as a function of halo central potential, $\phi_c$.  Lines are
best fit power laws $X \propto \phi_c^{\gamma}$, where X is $V_{\rm
cmax}^2$ and $\gamma = 1$ in the top panel and $M_{\rm vir}$ and $\gamma
= 1.5$ in the bottom.}
\label{fig:central_pot}
\end{figure}

The maximum circular velocity of a halo can be interpreted as a tracer
for the depth of halo central potential. It is more reliable for this
purpose than halo mass, which can be difficult to define, and
especially so for subhalos. Figure \ref{fig:central_pot} compares
$V_{\rm cmax}^{\rm 2}$ and $M_{\rm vir}$ as a function of $\phi_c$,
the gravitational potential at the point of peak density in the halo
(as identified from the smoothed particle distribution by the Denmax
routine in our halo finder \citep{Weller:04}). The solid lines in each
plot represent the best fit power-law. We obtain $V_{cmax}^2 \propto
\phi$ in the upper panel and $M_{vir}/M_* \propto \phi^{1.5}$ for the
lower. Although both plots show clear correlations, that of $V_{\rm
cmax}$ and $\phi_{c}$ (top panel) is clearly the tighter. To quantify
this, we adapt Equation \ref{eq:chi} to measure the goodness of fit of
the power-law in each case, obtaining $\sigma_{rel} = 0.08$ for the
$V_{cmax}^2 - \phi$ correlation and $\sigma_{rel} = 0.19$ for the
mass$- \phi$ correlation. This confirms that $V_{cmax}$ is the more
robust measure of halo central potential.

\subsection{Substructure}

We plot in Figure \ref{fig:mass_frac} the fraction of mass in
substructure as a function of halo mass.  The dashed line represents
the best fit power-law, $f_s = A(M_{vir} / M_*)^m$, with slope $0.44
\pm 0.06$ and amplitude $0.14 \pm 0.02$. Hence, higher mass halos tend
to have contain a higher fraction of their mass in susbtructure. This
is an important result, demonstrating that cluster-mass halos are
not self-similar. Several studies of small samples of halos selected
from lower resolution simulations have suggested the dark matter halos
are self-similar in terms of their subhalo populations: low mass halos
appear like `rescaled versions' of higher mass halos \citep{Moore:99a,
Ghigna:00, DeLucia:04, Reed:05}. With a {statistically significant
sample of well resolved halos, we find this not to be the
case. \cite{Gao:04}, who compare the subhalo populations of a small
sample of halos over a wide mass range, found a similar
result to the one presented here. \cite{vandenBosch:05} constructed a
semi-analytical model to calculate the subhalo mass function,
adjusting the free parameters to match the simulations of
\citet{Gao:04}, \citet{Tormen:04} and \citet{DeLucia:04}. They found
that the slope and normalization of the mass function to be dependent
on the ratio of halo mass to the characteristic non-linear mass
scale. \citet{Zentner:05} and \citet{Taylor:04a} have also proposed
models with similar properties, arguing that as higher mass halos have
formed more recently, they should contain a higher fraction of their
mass in substructure.

\begin{figure}
\plotone{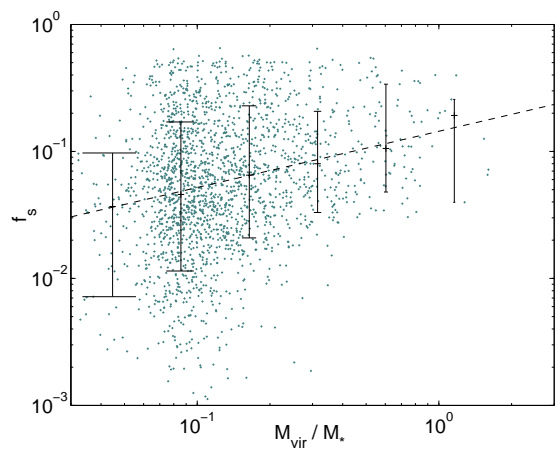}
\caption{The fraction of mass in substructure as a function of halo
mass. Points with errorbars represent the median values of $f_s$ in
logarithmically spaced mass bins. The dashed line is the best fit power
law, with slope $0.44 \pm 0.06$. We take $M_* = 8.0 \times 10^{14}
h^{-1} M_{\rm \sun}$.}
\label{fig:mass_frac}
\end{figure}

\subsection{Concentration}
\label{sec:mass_conc}

In their early studies, NFW postulated that the concentration of a
halo is an indicator of the mean density of the universe at the time
of its collapse \citep{Navarro:97}. Hence, to conform with the
current paradigm of hierarchical structure formation, halo
concentration ought to decrease with increasing mass.  Subsequent
studies \citep{Bullock:01b,Eke:01,Dolag:04} have
suggested analytic formulas predicting halo concentration as a
function of mass and redshift. \citet{Dolag:04} suggest a simple
fitting formula
\begin{equation}
c(M,z) = \frac{c_0}{1+z}\left(\frac{M}{M_*}\right)^{\alpha} \; ,
\label{eq:c}
\end{equation}
finding $\alpha = -0.102$ and $c_0 = 5.97$ for their $\Lambda$CDM
simulation, with $\alpha$ almost independent of cosmology. The
corresponding parameters obtained by \citet{Bullock:01b} are $\alpha =
-0.13$ and $c_0 = 5.60$. (It should be noted that both these studies
calculate the concentration by defining the virial radius as enclosing
an overdensity of $\Delta_c$ times the {\it mean background} density
rather than the the critical density as we do in this analysis. We
therefore have applied the appropriate correction to their results to
enable comparisons with our own.) Additionally, \citet{Avila-Reese:05}
found a logarithmic slope of $-0.14$ for $c(M_{vir})$, for the halos in
high density regions in their simulations. Analysing halo
concentrations at several redshifts, \citet{Bullock:01b},
\citet{Eke:01} and \cite{Navarro:04} all show that the concentration
parameter for a given mass is lower at higher redshifts. By following
the mass accretion history of simulated halos, \citet{Zhao:03b} and
\citet{Tasitsiomi:04} both find that halo concentration increases
during the slow accretion phase of halo formation, having remained
fairly constant during an initial merger period.

The median trend (points with error bars) in Figure
\ref{fig:mass_conc} clearly demonstrates the predicted
anti-correlation between concentration and virial mass at a constant
redshift for the halos in our sample. As in Section
\ref{sec:mass_vmax}, we have removed all halos with $f_s > 0.3$ as the
NFW profile provides a poor fit to many of these halos due to their
large quantity of substructure. The solid line represents the best fit
for the \citet{Dolag:04} fitting formula, with $\alpha = -0.12\pm0.03$
and $c_0 = 6.47\pm0.03$.  However, there is also a considerable amount
of scatter in the distribution of points making it difficult to
accurately determine the slope of the correlation. In their recent
study, \citet{Avila-Reese:05} showed that halo concentration is also
dependent on the local environmental density: halos in dense regions
are on average more concentrated and have higher central densities. As
demonstrated by \citet{Sheth:04}, halos in dense regions accumulate
their mass earlier than their counterparts in lower density
regions. Hence, low mass halos forming in low density regions will
have a lower concentration than predicted by the \citet{Dolag:04}
$M_{vir} - c$ relation. This can be seen in Figure
\ref{fig:mass_conc}--- there is a significant tail of low
concentration halos at lower masses.

\begin{figure}
\plotone{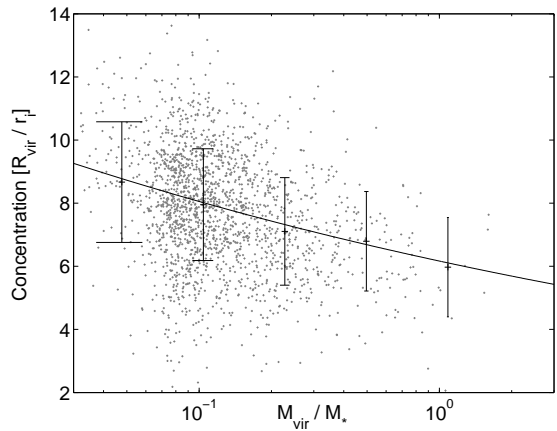}
\caption{Scatter plot of concentration as a function of virial mass,
having removed halos for which the NFW profile provides a poor fit due
to large amounts of substructure. The points with error bars are the
median concentration in logarithmic mass bins, the solid line is best
fit of Eq. \ref{eq:c} to the data ($\alpha = -0.12$ and $c_0 =
6.47$).}
\label{fig:mass_conc}
\end{figure}

\subsection{Morphology}

In order to investigate how halo morphology is dependent on mass, we
plot in the upper of Figure \ref{fig:mass_shape} the minor to major
axis ratio versus halo mass. The points with errors bars mark the
median trend in log-linear mass bins. The solid line indicates the
best fit power law, $c/a \propto (M_{vir} / M_*)^\delta$, with $\delta
= -0.049 \pm 0.007$. The slope, $\delta$, that we obtain is in very
good agreement with a comprehensive recent study of halo morphology by
\citet{Allgood:05}, who measure $\delta = -0.050 \pm 0.003$ over a
wide range of halo masses. In another recent paper, \citet{Paz:05}
measure a steeper relationship, with $\delta = -0.056 \pm
0.003$. \citet{Jing:02}, who analyse a similar halo mass range to
ours, find a steeper slope still. As pointed out by \cite{Allgood:05},
who perform an in-depth comparison of halo morphologies from different
simulations and measuring techniques, this is probably due to the
approach that \citet{Jing:02} adopt of measuring the shapes defined by
isodensity contours, thus ignoring the mass distribution in the
central regions of their halos. Essentially, this suppresses the
effect on halo shape of large substructures near the halo
core.  Regardless of the shape measuring technique used,
it is clear that in general halos become less spherical with mass.

In the lower panel of Figure \ref{fig:mass_shape} we plot halo
triaxiality (as defined in Equation \ref{eq:T}) versus mass. As
implied by the size of the error bars, there is an enormous amount of
scatter. However, there is a preference for high mass halos to be more
prolate. Therefore, the apparent trend is for dynamically `younger'
halos (larger $ M_{vir} / M_*$) to be more prolate. Low mass halos are
older than their higher mass counterparts; they have had longer to
dynamically relax and obtain a spherical morphology than the higher
mass, more prolate halos. Furthermore, higher mass halos contain a
higher fraction of their mass in substructure, which also influences
their shape. This is investigated further in the following Section.

\begin{figure}
\plotone{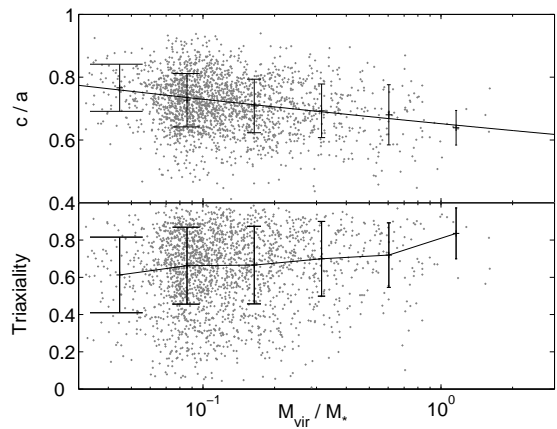}
\caption{Scatter plot of halo virial mass, $M_{\rm vir}$, versus the
  minor to major axis ratio ({\it top}) and halo triaxiality ({\it
  bottom}). The solid line in the top panel represents the best fit
  power law, with slope $ -0.049 \pm 0.007$ (see text). Points with
  error bars denote median value and standard deviation in log-linear
  mass bins. The solid line in the bottom panel is the median trend
  line. The plots show that higher mass halos are less spherical
  (decreasing c/a) and more prolate (T approaching unity).}
\label{fig:mass_shape}
\end{figure}

\section{The impact of substructure on halo properties}
\label{sec:impactsub}

We now investigate the impact of substructure on the physical
properties of the dark matter halos.  Specifically, the purpose is to
determine how the distribution of subhalos effect the morphology,
concentration and spin of a cluster. All the results presented in this
Section are for the `virialised' halos in our sample only.

\subsection{Mass distributions and concentration}

In the previous Section we found that halo concentration--- an
indicator of the formation epoch of a halo--- decreases with halo
mass. We find there to be an even tighter anti-correlation between
halo concentration and the total fraction of mass in substructure:
halos with more substructure have lower concentrations.  This is
evidence of the continuous effect of the tidal stripping of
subhalos. Halos with large amounts of substructure formed more
recently, when the background density of the universe--- and therefore
central densities and concentrations--- are lower. The less dense
outer parts of infalling subhalos are stripped off in the outer
regions of the halo, and the denser core spirals into the
central region due to dynamical friction, thus increasing the
concentration of the halo once the subhalo is fully digested.

\subsection{Contribution of substructure to halo morphology}

We have already shown that substructures lying just outside of the
virial radius have a measurable effect on the overall distribution for
each axial ratio. Figure \ref{fig:fracT} is a scatter plot of the
fraction of mass in substructure against triaxiality. Again, there is
an appreciable amount of scatter, although the apparent trend is for
substructure-rich halos to be more prolate.  To further quantify the
effect of substructure, we recalculate the c/a axis ratio for each
halo, but having first removed all particles that belong to
substructure, leaving particles that belong only to the `mother' or
`host' halo. We also repeat this exercise, but removing only the most
massive subhalo in the cluster and retaining the rest.  Figure
\ref{fig:shape_nosubs} shows the change in the minor/major axis ratio
when these two sets of particles are removed.  These distributions
indicate that the particles belonging to the mother halo only are more
spherically distributed than the rest of the halo mass.  Therefore,
the effect of substructure is to make cluster size halos appear more
prolate. The dashed histogram demonstrates that much of the impact of
substructure on the minor/major axis ratio is due to the most massive
subhalo in each halo. Smaller substructures, regardless of their
position in the cluster, appear to have less influence.

\begin{figure}
\plotone{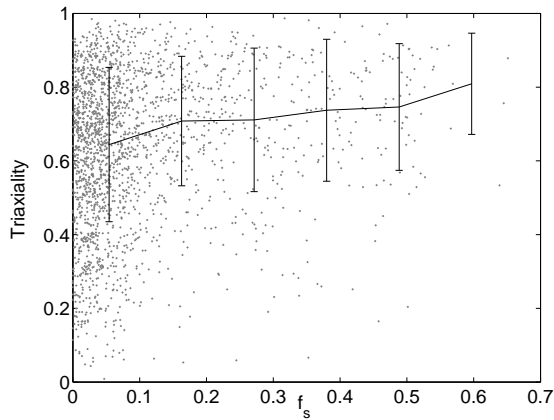}
\caption{Scatter plot of the fraction of mass in substructure against
halo triaxiality. Error bars represent the standard deviation in each
bin. There appears to be a slight trend for triaxiality to increase
with $f_s$, although the errors on each point are large. This
indicates a tendency for halos to become more prolate as the fraction
of their mass in substructure increases. }
\label{fig:fracT}
\end{figure}
\begin{figure}
\plotone{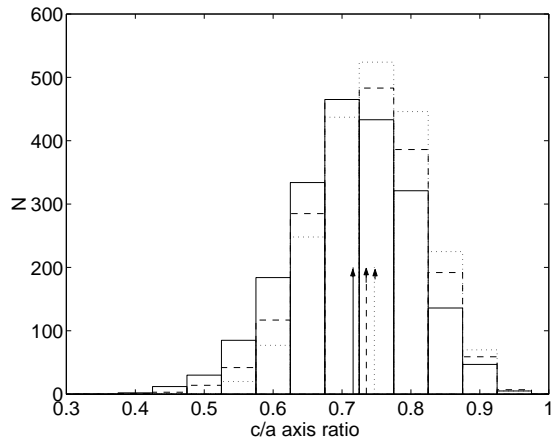}
\caption{Histograms of the c/a axis ratio having included (solid) and
omitted (dotted) the substructure whilst determining $I_{ij}$ for each
halo. Also plotted is the distribution having removed only the most
massive subhalo (dashed). Arrows denote the median values of each
distribution. The dotted histogram shows that when the substructure is
discarded, thus leaving only the mother halo, the shape becomes more
spherical (c/a closer to unity). 
It is clear that the largest subhalo is
significant in determining cluster morphology: removing these
results in an obvious shift in the overall c/a distribution.}
\label{fig:shape_nosubs}
\end{figure}

\subsection{Effect of substructure on halo spin}

Figure \ref{fig:fraclam} shows the relation between halo angular
momentum and the fraction of mass in substructure.  The median value
of $\lambda$ increases from 0.030 for halos with $f_s < 0.2$, to 0.051
for those with $f_s > 0.2$. This indicates that halos with higher
$f_s$--- those more likely to have undergone major mergers--- have had
their angular momentum boosted through the acquisition of smaller
halos.

\begin{figure}
\plotone{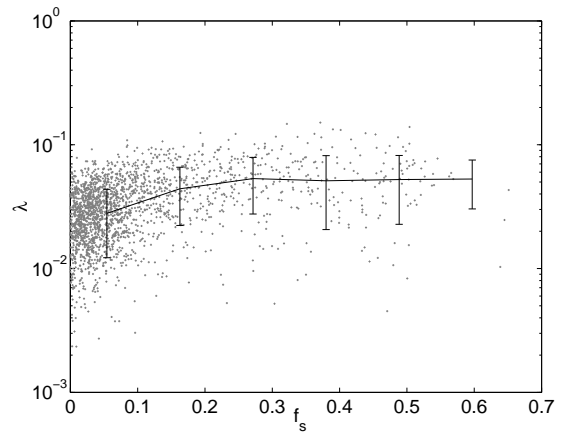}
\caption{Spin parameter
$\lambda$ as a function of the fraction of mass in
substructure, $f_s$. The median trend line shows there to be, on
average, a small increase in halo spin, as the fraction of halo mass
in substructure increases.}
\label{fig:fraclam}
\end{figure}

To investigate this further, we now consider the contribution to 
the total spin by the mother halo only and by the substructure.
To calculate these quantities we
define a basic unit angular momentum per unit mass
\begin{equation}
[j]  \equiv \left[ 2GR_{\rm vir}M_{\rm vir} \right]^{\frac{1}{2}} \; ,
\end{equation}
where $R_{\rm vir}$ and $M_{\rm vir}$ are the total halo energy and mass
respectively.
We then calculate the angular momentum per unit mass directly for
particles belonging solely to the mother halo ($\lambda_{m}$) and for
those belonging to substructure ($\lambda_{s}$) in terms of [j]:
\begin{equation}
\lambda_{m} \equiv \frac{(J_{m}/M_{m})}{[j]} \; , \; \; \; \;
\lambda_{s} \equiv \frac{(J_{s}/M_{s})}{[j]} \; ,
\label{eq:unitspin}
\end{equation}
where $J_{m}$ and $J_{s}$ are calculated using particles belonging
only to the mother halo or substructure, respectively. 
Note if $f_s=0$, then $\lambda_{s}$ is also zero and $\lambda_{m}$
returns to our original definition of $\lambda$.

Figure \ref{fig:sublam} shows the overall distributions obtained for
$\lambda_{s}$ and $\lambda_{m}$ compared to the original
distribution. We have also plotted the best fit log-normal curves
(Equation \ref{eq:lognorm}) where $\lambda_{0m}$ = 0.029 and
$\sigma_{m}$ = 0.608 for the mother halo, and $\lambda_{0s}$ = 0.144
and $\sigma_{s}$ = 0.863 for the substructure. These results clearly
show that particles belonging to substructure have a far greater
average angular momentum per unit mass than those belonging solely to
the mother halo. Accreted particles will fall into a halo in a
generally spherically symmetric manner, and will therefore induce only
a smaller change to the total angular momentum compared to the capture
of a subhalo--- an injection of mass from one particular
direction. Importantly, in the next Section we find that subhalos tend
to be found in a prograde orbit with respect to the overall angular
momentum of the mother halo, implying a preferential direction of
infall. However, there is also a selection effect here. Subhalos
adopting plunging orbits directly towards the cluster center of mass,
and therefore possessing a lower orbital angular momentum, will be
more rapidly destroyed through their exposure to tidal forces from the
halo core. This introduces a bias favoring the survival of subhalos in
more spherical orbits.

Overall, we expect the velocity distribution of captured dwarf halos
to be fairly broad--- those that form near to their eventual parents
will have a lower velocity relative to the mother than those that
have formed further away and have thus gained more kinetic energy as
they fell into the potential well of their future host.  This effect
explains why $\sigma_{s}$, the dispersion in $\lambda_{s}$, is so
great.

\begin{figure}
\plotone{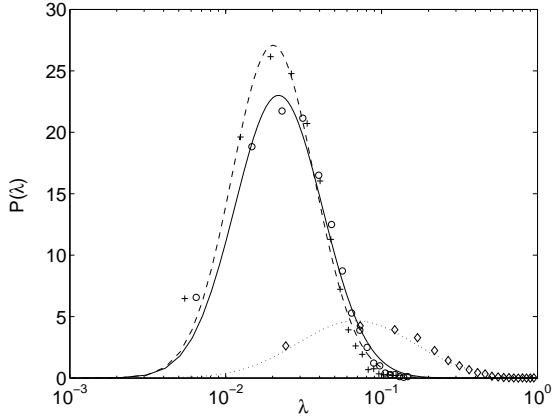}
\caption{Distributions of $\lambda$ (circles), $\lambda_m$ (crosses)
  and $\lambda_s$ (diamonds), and the respective fits to a log-normal
  distribution (solid, dashed, and dotted lines; see Equation
  \ref{eq:lognorm}). $\lambda_m$ and $\lambda_s$ are calculated using
  particles belonging either only to the mother halo or substructure,
  respectively. $\lambda$ is plotted on a log scale due to the much
  greater angular momentum per unit mass (see Equation
  \ref{eq:unitspin}) of particles belonging to substructure.  The fit
  parameters are $\lambda_{0m}$ = 0.029, $\sigma_{m}$ = 0.608 for the
  mother halos only, and $\lambda_{0s}$ = 0.144 and $\sigma_{s}$ =
  0.863 for the substructure only.}
\label{fig:sublam}
\end{figure}

In order to investigate the effect of substructure on the spin of the
mother halo, we plot $\lambda_{s}$ and $\lambda_{m}$ against $f_{s}$
in Figure \ref{fig:fraclamsm}. In the equivalent plot for the total
halo spin, we found there to be a correlation between $\lambda$ and
$f_s$ (Figure \ref{fig:fraclam}). We find a similar result for the
mother halo spin, but a weak anti-correlation between substructure
spin and $f_s$. At first glance this may seem strange, but it is
merely due to the fact that halos with a large $f_s$ tend to contain
at least one very massive piece of substructure. Being located near
the halo center, the subhalo does not contribute as much angular
momentum as several smaller and more distant subhalos would, resulting
in a decreased value of $\lambda_{s}$.

\begin{figure}
\plotone{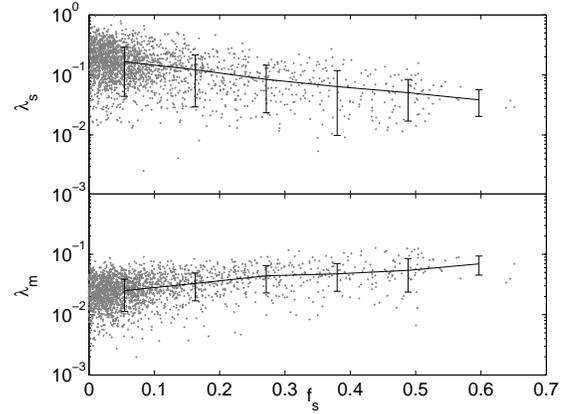}
\caption{({\it top}) Similar to Figure \ref{fig:fraclam}, except for
$\lambda$ is calculated only using particles that belong to
substructure in each halo. The median trend-line shows a weak decrease
of $\lambda_s$ with $f_s$--- the opposite to the trend demonstrated in
\ref{fig:fraclam}, for total halo spin. ({\it bottom}) Same as above
except for $\lambda$ calculated only using particles that belong 
to the mother halo.  Median trend-line shows an increase of
$\lambda_m$ with $f_s$. Combined, these plots demonstrate that the
increase of total halo spin with $f_s$ is due only to the
increase of $\lambda_m$ with $f_s$.}
\label{fig:fraclamsm}
\end{figure}

The correlation between $\lambda_{m}$ and $f_s$ is a significant
result. As $\lambda_{s}$ decreases with increasing $f_s$, the overall
correlation between $\lambda$ and $f_s$ (see Figure \ref{fig:fraclam})
must be entirely due to the increase in $\lambda_{m}$. We are
therefore seeing evidence of the transfer of angular momentum from
captured subhalos, inducing a `spin-up' of the mother (we
re-emphasize that the mother is that largest structure in each
halo, consisting of all the particles that do not belong to any
substructures). There are two possible processes by which this can
occur.  In the first scenario, a recently absorbed dwarf halo is
slowed by gravitational dynamical friction as it penetrates into the
cluster. This results in a transfer of energy and angular momentum to
the surrounding particles, whilst slowing the captured halo into a
more stable orbit (see \citet{Zhao:04} for a detailed study of this
process). Hence, even mother halos that host a single massive subhalo
near the halo center will have a greater spin due to the transfered
angular momentum as the subhalo migrates towards the
center. \citet{Peirani:04} and \citet{Vitvitska:02} directly measure
this effect by comparing the radial angular momentum distribution
before and after a merger episode (see Figure 6 in
\citet{Peirani:04}). In general they find that the maximum specific
angular momentum increases sharply in the aftermath of the merger.
This process is also seen indirectly by \citet{Gardner:01} in his
simulations--- halos that experienced a recent merger have a larger
spin than those that have grown purely through gradual accretion.

The second process by which substructure can transfer angular
momentum to the host halo is through the tidal stripping of their
outer regions as they orbit around the halo core.  As Figure
\ref{fig:sublam} shows, particles belonging to substructure have a
greater angular momentum per unit mass than those that belong to the
host halo. Therefore, as particles are stripped from substructure and
become part of the mother halo, the mean angular momentum of the
mother increases.

\subsection{Alignment of host and subhalo orbital angular momentum}

In the preceding Section we demonstrated that clusters with a larger
fraction of their mass in substructure have a greater spin. This is
due to the higher average orbital angular momentum per unit mass of
subhalos compared to the mother halo. However, for this to be true the
majority of subhalos must have a prograde orbit with respect to the
mother halo particles, i.e. the angular momentum vector for the
substructure must be well aligned with that of the host. In Figure
\ref{fig:jangles} we investigate this by plotting the distribution of
the relative orientations of the angular momentum vectors of
substructure particles ($\mathbf{J}_{s}$), mother halo particles
($\mathbf{J}_{M}$) and the cluster as a whole ($\mathbf{J}_{tot} =
\mathbf{J}_{s} + \mathbf{J}_{M}$) for each halo in our sample. The
middle panel ($\mathbf{J}_{s} \cdot \mathbf{J}_{M}$) is of the most
interest. It clearly demonstrates that there is a tendency for
$\mathbf{J}_{s}$ and $\mathbf{J}_{M}$ to be closely aligned. Hence, it
appears that subhalos do tend to orbit in the same direction as the
host halo particles.

It has been demonstrated both through observations \citep{Plionis:02}
and in simulations \citep{Colberg:05,Knebe:04,Faltenbacher:02} that
subhalos preferentially fall into clusters along the filamentary
structures surrounding them. \citet{Tormen:97} and \citet{Knebe:04}
both conclude that infall along filaments account for both the cluster
shape and velocity structure. The correlation between the orientation
of the angular momenta of mother halo and substructure particles is
just a consequence of the same process responsible for the spinning up
of a collapsing halo. The angular momentum of a halo is acquired
through the influence of the tidal fields of neighboring structures
acting on the infalling particles and subhalos during halo formation
\citep[see, for example][and references therein]{White:84}, the influx
of which is primarily from the high density regions that eventually
form filamentary structures. Once captured, the outer region of a
subhalo is tidally stripped, and the dense remnant spirals into and is
eventually consumed by the halo core. Thus, the mother accumulates all
the orbital angular momentum of the subhalo. Over time, as subhalos
are captured and absorbed, the the spin of the mother halo gradually
increases in this manner. Hence, the measured alignment between the
angular momentum of the mother halo and subhalo particles is merely
due to the temporal auto-correlation of the angular momenta of matter
accumulated by the cluster.

Additionally, a smaller halo on a prograde orbit around a cluster will
gravitationally perturb nearby particles within the virial radius,
causing a slight extension of the halo shape towards it
\citep{Tremaine:84}. This in turn increases the attraction of the
cluster on the halo, thus improving the likelihood that it will be
captured.  Halos that pass the cluster in retrograde orbits will not
be able to perturb the cluster to as great an effect. Hence, there is
a slightly greater capture probability for a halo that has a lower
{\it relative} velocity to the local cluster particles, whether it
arrives from a filament or not.

\begin{figure}
\plotone{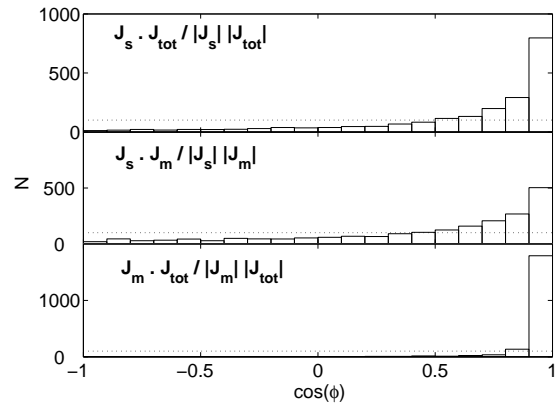}
\caption{Distributions of the relative orientation of the angular
  momentum vectors for the entire cluster and the substructure (top
  panel), the substructure and the mother halo (middle panel) and the
  mother halo and entire cluster (bottom panel). The middle plot
  demonstrates that there is reasonably good alignment (to within
  $25^o$) of the angular momentum of the mother halo and subhalos.}
\label{fig:jangles}
\end{figure}

\subsection{Alignment of angular momentum and halo axes}

For a rotating body, it is expected that the direction of the angular
momentum vector $\mathbf{J}$ will align with one of the principle axis
vectors.  The most stable configuration is when $\mathbf{J}$ is
parallel to the minor axis $\mathbf{c}$. Figure \ref{fig:jdotaxis}
displays the distribution of the cosine of the angle $\phi$ between
the angular momentum vector and the major, intermediate and minor axes
for all the halos in our sample. There is clearly a tendency for the
angular momentum vector to be aligned with the minor axis of the
halos. Correspondingly, the major axis is far more likely to be nearly
perpendicular to $\mathbf{J}$. This result was also found by
\citet{Barnes:87}, \citet{Warren:92a}, \citet{Bailin:04},
\citet{Allgood:05} and \citet{Avila-Reese:05}.

The intermediate axis alignment is interesting: it appears to show no
preferred alignment with $\mathbf{J}$, having an almost uniform
distribution over all cos$\phi$. We have investigated whether this may
be due to the difficulties in determining the minor axis when c/b is
close to unity. However, we found this not to be the case--- the axis
ratios are widely distributed for these halos. Hence, a minority of
halos are aligned so that their intermediate axes is parallel to the
direction of the angular momentum vector--- an unstable configuration
for a solid rotating body. In Figure \ref{fig:angs_axis} we break down
the range of $\lambda$ into four equally sized sets. The purpose of
this is to investigate whether the oddly aligned halos are those with
the least (normalized) angular momentum. It is clear that as $\lambda$
decreases the distribution of orientations becomes more evenly
spread. Therefore, the orientation of a halo is therefore dependant on
the magnitude of its spin. The greater the angular momentum of a halo,
the more likely it is to be found rotating around the minor principle
axis, and the less likely around the major axis. Except for the most
rapidly rotating halos, which maintain the expected orientation, the
results for the intermediate axis are inconclusive. The seemingly
random distribution of alignments is maintained for slowly rotating
halos.

\begin{figure}
\plotone{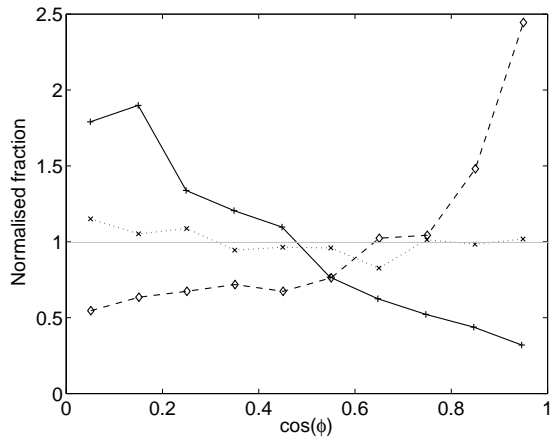}
\caption{Distribution of the cosine of the angle $\phi$ between the
  angular momentum vector and the major (solid), intermediate (dotted)
  and minor (dashed) axes. The plot is normalized so that a uniform
  distribution would have value of unity on the vertical axis (grayed
  horizontal line).}
\label{fig:jdotaxis}
\end{figure}
\begin{figure}
\plotone{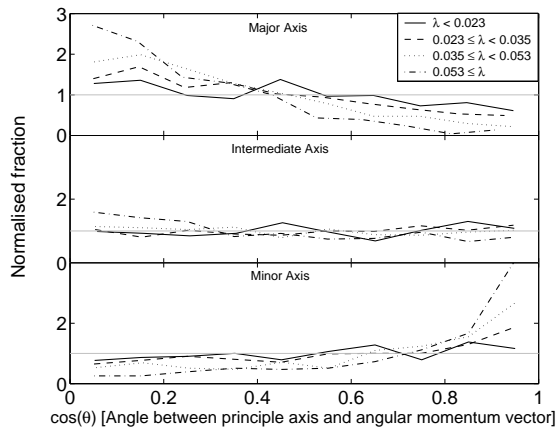}
\caption{The distribution of the cosine of the angle between the
  angular momentum vector and the major (top panel), intermediate
  (middle), and minor (bottom) axis for halos grouped into ranges of
  total halo spin, $\lambda$. The plot demonstrates an improvement in
  alignment as $\lambda$ increases. Rapidly spinning halos have a
  tendency for $\mathbf{J}$ to be more closely aligned with the minor
  axis. Each plot is normalized so that a uniform distribution would
  have value of unity on the vertical axis (grayed horizontal lines).}
\label{fig:angs_axis}
\end{figure}

\subsection{Correlation between $\lambda$ and triaxiality}

We now consider whether the \citet{Peebles:69} relation between the
ratio of the major and minor principle axes and the cluster spin
(Equation \ref{eq:polytrope}) holds for the dark matter halos in our
simulation. As we determined in Section \ref{sec:morphology}, most of
the halos tend to be prolate rather than oblate.  Here we investigate
whether there is an increased departure from a spherical morphology
(i.e. a reduction of the axis ratio c/a) as $\lambda$ increases.

There are at least two effects that could influence this
correlation. The first of these is the possibility of a significant
misalignment between the minor principle axis and the angular momentum
vector. Equation \ref{eq:polytrope} assumes that the object is
rotating around its minor principle axis. In those halos for which
there is a significant misalignment between the minor axis and the
angular momentum vector, the effect of halo spin on shape will be
unpredictable. However, as many of the halos in our sample are
prolate, with a mean c/b axis ratio near 0.9, we include only those
halos for which the angle between the major axis and the angular
momentum vector is greater than or equal to $70^o$.  A second possible
source of uncertainty will be the effects of substructure on the halo
shape. Figure \ref{fig:shape_nosubs} showed the principle axis ratios
using particles belonging only to the mother halo (i.e. having removed
the subhalos) finding it to be more spherical than the entire halo.
Equation \ref{eq:polytrope} applies to a cylindrically symmetric and
uniformly rotating polytrope, and therefore will certainly not hold
when there is significant substructure in the cluster halo. Hence we
discard all particles belonging to substructure before calculating the
c/a axis ratio for each halo, and we use $\lambda_m$--- the spin
calculated using the mother halo particles only--- rather than
$\lambda$.

Even having accounted for these possible sources of discrepancy, we
find that there is no correlation between the spin and the overall
shape of the halos. It is therefore clear that other processes are
responsible for determining halo morphology. The mean value we obtain
for $\lambda_0$ is 0.033. Therefore, cluster mass halos cannot support
themselves through rotation alone, it is instead through their
velocity dispersion that they are able to do so. Hence, spin is
largely irrelevant in determining shape even though we find a strong
correlation between the angular momentum vector and the minor
principle axis. We have also shown that the angular momentum of dark
matter halos is increased as subhalos are accumulated, which also
influences halo shape. Furthermore, subhalos tend to be found in
prograde orbits, partly because of a preferential direction of infall
from filaments surround the halo. It appears that the orientation of
dark matter clusters can be explained by the specific merging
directions of smaller halos that they capture, during their formative
years.

\section{Discussion and Conclusion}

In this study, we set out to investigate the overall physical
characteristics of a sample of over 2000 massive dark matter cluster
halos identified in a 320 $h^{-1}$Mpc, 1024$^3$ particle simulation.
Using the algorithm described in \citet{Weller:04}, all particles
belonging to substructure in each cluster were identified, down to a
minimum subhalo mass of 30 particles, or $\sim 10^{11} h^{-1} M_{\rm
\sun}$.  We then applied two further criteria to our sample. First, to
ensure that all the halos in our dataset were well resolved--- thus
ensuring that no substructure has been washed out by the so-called
`over-merging' problem--- we set a minimum cluster halo mass of 10,000
particles, leaving us with an overall sample of 2159 halos. Second, we
identified and tagged those halos not yet in a state of dynamical
equilibrium. The aim of the second criterion is to enable separate
analyses of those halos that are virialized and those that are still
at an earlier stage of their formation.

In order to implement this second criterion, we defined a measure of
virialization, $\beta$ (Equation \ref{eq:beta}), based on the virial
theorem and including a necessary term to compensate for the surface
pressure at the virial radius of each halo. Using this technique, we
demonstrated quantitavely that higher mass, substructure-rich halos
are less virialized. We removed from our sample any halo with
$\beta<-0.2$ (i.e. those still collapsing) -- which make up roughly
3.4$\%$ of the entire sample -- tagging them as `unvirialized'. In
total, our sample contained 2085 virialized halos, each containing
between 10,000 and 500,000 particles. Such a large number of well
resolved halos allowed us to determine their physical properties in a
statistically significant way.  

We then constructed a catalogue of their physical properties,
including concentration, morphology, angular momentum, maximum
circular velocity, and the fraction of mass in substructures. The main
questions motivating our analysis were: how are the halo properties
distributed, both in our main halo sample and in those that we have
tagged as `unvirialized'?  Do halo properties depend on mass? And how
are they influenced by substructure?  These questions can only be
answered if one analyses a sample of considerable size with no prior
selection of specific halos to resimiulate from a lower resolution
simulation. The distributions obtained in this work are drawn from a
large enough sample (>2000) to address these questions in a
statistical meaningful way.

The halos in our virialized sample contained a median value of 5.6\%
of their mass in substructure, with a tail to large substructure
fractions. On average, they tended to have a prolate morphology
and a mean minor to major principle axis ratio of $c/a = 0.707 \pm
0.095$. The unvirialized halos were more substructure-rich, with a
median value of 19.6\% of their mass in substructure. They also tended
to be less spherical, have a higher spin and a lower concentration
than their dynamically relaxed counterparts. 

We have clearly demonstrated that the physical properties of halos are
dependant on their mass and therefore their age. The median fraction
of mass contained in substructure increases with halo mass as $f_s
\propto (M_{\rm vir} / M_*)^m$, where $m = 0.44 \pm 0.06$.  This
supports the prediction made by \citet{Zentner:05} that as higher mass
halos have formed more recently than lower mass halos, they should
contain more substructure as they have had less time to disrupt their
internal subhalo population. One direct consequence of this is that
higher mass halos appear more prolate: we find $c/a \propto (M_{vir} /
M_*)^\delta$, where $\delta = -0.049 \pm 0.007$. As expected, halo
concentration decreases as mass increases, with $c(M) \propto (M /
M_*)^\alpha$ with $\alpha$ = -0.12 $\pm$ 0.03, agreeing well with the
slopes measured by \citet{Dolag:04} and \citet{Bullock:01b}.

Halos with greater amounts of substructure tend also to have a higher
spin; the angular momentum per unit mass of subhalos is far greater
than that of the mother halo. We also find evidence of the transfer of
angular momentum to the mother halo (which consists of all the
particles in the halo that do not belong to substructures) from
captured subhalos, via dynamical friction and the tidal stripping of
their outer regions. Furthermore, the orbital angular momentum of
subhalos is typically well aligned with that of their host. We propose
that this is mainly due to the process by which a halo accumulates
angular momentum from captured subhalos, which, due to large-scale
tidal fields, have a preferential direction of infall from the
surrounding filaments \citep[see, for example][]{Colberg:05}. The
angular momentum of consumed subhalos is added to that of the mother
halo, so that over time it will begin to rotate in the same direction
to which the subhalos are `injected', explaining the observed
alignment. Additionally, we find that the orientation of the angular
momentum vector of halo with respect its morphology is dependant its
magnitidue. Rapidly spinning halos tend to align with the minor
principle axis; halos with very low values of $\lambda$ are have a
higher probabiliy of being aligned with the intermediate or major
axes. However, angular momentum is not responsible for determining
halo shape. Indeed, we concluded that the correlation between angular
momentum and the minor principle axis is due to a common influence on
both properties--- the preferential merging direction of subhalos
along halo filaments.

\acknowledgements 
JW was supported by the DOE and the NASA grant NAG
5-10842 at Fermilab.  This work was supported by a grant of
supercomputing time (grant number MCA04N002P) from the National Center
for Supercomputing Applications.  This research also used
computational facilities supported by NSF grant AST-0216105. LS thanks
Antonio Vale for helpful discussions and PPARC for financial support.

\end{document}